\documentstyle [12pt] {article}
\begin{document}
\setcounter{page}{1}
\def\theequation{\arabic{section}.\arabic{equation}}
\def\theequation{\thesection.\arabic{equation}}
\setcounter{section}{0}

\title{Solar proton burning,\\  photon and anti--neutrino disintegration of
the deuteron
in the relativistic field theory model of the deuteron}

\author{A. N. Ivanov~\thanks{E--mail: ivanov@kph.tuwien.ac.at,
Tel.: +43--1--58801--14261, Fax: +43--1--5864203}~${\textstyle ^\ddagger}$,
H. Oberhummer~\thanks{E--mail: ohu@kph.tuwien.ac.at, Tel.:
+43--1--58801--14251, Fax: +43--1--5864203} ,
N. I. Troitskaya~\thanks{Permanent Address:
State Technical University, Department of Nuclear
Physics, 195251 St. Petersburg, Russian Federation} ,
M. Faber~\thanks{E--mail: faber@kph.tuwien.ac.at,
Tel.: +43--1--58801--14261, Fax: +43--1--5864203}}

\date{}

\maketitle

\begin{center}
{\it Institut f\"ur Kernphysik, Technische Universit\"at Wien,\\
Wiedner Hauptstr. 8-10, A-1040 Vienna, Austria}
\end{center}

\vskip1.0truecm
\begin{center}
\begin{abstract}
The relativistic field theory model of the deuteron (RFMD) is applied to
the calculation of the astrophysical factor $S_{\rm pp}(0)$ for the process
of the solar proton burning p + p $\to$ D + e$^+$ + $\nu_{\rm e}$ and the
cross sections for the disintegration of the deuteron by photons $\gamma$ +
D $\to$ n + p and anti--neutrinos $\bar{\nu}_{\rm e}$ + D $\to$ e$^+$ + n +
n. Our theoretical value of the astrophysical factor $S_{\rm pp}(0) =
4.02\times\,10^{-25}\,{\rm MeV\,b}$ agrees with the classical result
obtained by Bahcall and Kamionkowski $S_{\rm pp}(0) =
3.89\times\,10^{-25}\,{\rm MeV\,b}$ in the potential model approach (PMA).
The cross sections for  the disintegration of the deuteron by photons and
anti--neutrinos calculated near thresholds are in good agreement
with the PMA. An extrapolation of the cross sections for energies far from
thresholds is suggested and related to the inclusion of form factors
describing spatial smearing of the deuteron and the NN system. The
extrapolated cross section for the disintegration of the deuteron by
anti--neutrinos agrees with that calculated in the PMA in the
anti--neutrino energy region from threshold up to $E_{\bar{\nu}_{\rm e}} =
10\,{\rm MeV}$. The extrapolated cross section averaged over the reactor
anti--neutrino energy spectrum is obtained in agreement with the
experimental data. It is shown that the RFMD enables to describe elastic
low--energy NN scattering in accordance with low--energy nuclear
phenomenology.
\end{abstract}
\end{center}

\begin{center}
\end{center}

\newpage

\section{Introduction}
\setcounter{equation}{0}

\hspace{0.2in} The relativistic field theory model of the deuteron (RFMD)
formulated in Refs.~[1,2] has been applied to the calculation of the
reaction rate of the neutron--proton radiative capture n + p $\to$ D +
$\gamma$ and the astrophysical factor $S_{\rm pp}(0)$ of the solar proton
burning p + p $\to$ D + e$^+$ + $\nu_{\rm e}$ [2]. Some mistakes, which had
been made for the first calculation [2], have been then partly, mainly for
the reaction rate of the neutron--proton radiative capture, corrected in
Ref.~[3]. However, our conclusion concerning the value of the astrophysical
factor $S_{\rm pp}(0)$ [3] is still erroneous.

In this paper we would like to amend our results obtained in Refs.~[1--3]
and to apply the RFMD to the calculation of the cross sections for the
disintegration of the deuteron by photons $\gamma$ + D $\to$ n + p and
anti--neutrinos  $\bar{\nu}_{\rm e}$ + D $\to$ e$^+$ + n + n. The reaction
of the disintegration of the deuteron by anti--neutrinos $\bar{\nu}_{\rm
e}$ + D $\to$ e$^+$ + n + n is caused by the charged weak current and
valued, in the sense of charge independence of the weakinteraction
strength, to be equivalent to the observation of the reaction of the solar
proton burning p + p $\to$ D + e$^+$ + $\nu_{\rm e}$ in the terrestrial
laboratories [4]. Experimentally the reaction of the disintegration of the
deuteron by anti--neutrinos $\bar{\nu}_{\rm e}$ + D $\to$ e$^+$ + n + n
induces itself by reactor anti--neutrinos with an equilibrium energy
spectrum [5,6]. Therefore, experimental data  on the reaction
$\bar{\nu}_{\rm e}$ + D $\to$ e$^+$ + n + n are given in the form of the
cross section averaged over the reactor anti--neutrino energy spectrum
[4,7--9].

We show that all processes under consideration can be described in the RFMD
in agreement with the potential model approach (PMA) in spite of completely
different dynamics of strong low--energy nuclear interactions. Indeed, in
the RFMD [1,2] the physical deuteron appears through long--wavelength
vacuum fluctuations of the proton and the neutron field in the one--nucleon
loop approximation. In terms of one--nucleon loop exchanges we describe in
the RFMD a non--trivial wave function of the relative movement of the
nucleons inside the physical deuteron. Therefore, the physical deuteron
couples to nucleons and other particles only through one--nucleon loop
exchanges.

In order to couple to the deuteron through the one--nucleon loop exchange
the nucleons should pass through intermediate interactions providing
low--energy transitions N + N $\to$ N + N. In a quantum field theory
approach such interactions should be induced by  meson exchanges. As the
nucleons couple at low energies, the main contribution should come from the
one--pion exchange. The contributions of heavier meson exchanges can be
taken into account effectively by integrating them out.

Since in the reactions  p + p $\to$ D + e$^+$ + $\nu_{\rm e}$, n + p $\to$
D + $\gamma$, $\gamma$+ D $\to$ n + p and $\bar{\nu}_{\rm e}$ + D $\to$
e$^+$ + n + n the nucleons couple in the ${^1}{\rm S}_0$--state, the
low--energy transitions N + N $\to$ N + N can be described by the effective
local four--nucleon interactions [1,2]:
\begin{eqnarray}\label{label1.1}
&&{\cal L}^{\rm NN \to NN}_{\rm eff}(x)=G_{\rm \pi
NN}\,\{[\bar{n}(x)\gamma_{\mu}
\gamma^5 p^c(x)][\bar{p^c}(x)\gamma^{\mu}\gamma^5 n(x)]\nonumber\\
&&+\frac{1}{2}\,
[\bar{n}(x)\gamma_{\mu} \gamma^5 n^c(x)][\bar{n^c}(x)\gamma^{\mu}\gamma^5
n(x)] +
\frac{1}{2}\,[\bar{p}(x)\gamma_{\mu} \gamma^5 p^c(x)]
[\bar{p^c}(x)\gamma^{\mu}\gamma^5 p(x)]\nonumber\\
&&+ (\gamma_{\mu}\gamma^5 \otimes \gamma^{\mu}\gamma^5 \to \gamma^5 \otimes
\gamma^5)\},
\end{eqnarray}
where $n(x)$ and $p(x)$ are the operators of the neutron and the proton
interpolating fields, $n^c(x) = C \bar{n}^T(x)$, etc., then $C$ is a charge
conjugation matrix and $T$ is a transposition.
The effective coupling constant $G_{\rm \pi NN}$ is defined by
\begin{eqnarray}\label{label1.2}
G_{\rm \pi NN} = \frac{g^2_{\rm \pi NN}}{4M^2_{\pi}} - \frac{2\pi a_{\rm
np}}{M_{\rm N}} = 3.27\times 10^{-3}\,{\rm MeV}^{-2},
\end{eqnarray}
where $g_{\rm \pi NN}= 13.4$ is the coupling constant of the ${\rm \pi NN}$
interaction, $M_{\pi}=135\,{\rm MeV}$ is the pion mass, $M_{\rm p} = M_{\rm
n} = M_{\rm N} = 940\,{\rm MeV}$ is the mass of the proton and the neutron
neglecting the electromagnetic mass difference, which is taken into account
only for the calculation of the phase volumes of the final states of the
reactions p + p $\to$ D + e$^+$ + $\nu_{\rm e}$ and $\bar{\nu}_{\rm e}$ + D
$\to$ e$^+$ + n + n, and $a_{\rm np} = (-23.748\pm 0.010)\,{\rm fm}$ is the
S--wave scattering length of the np scattering in the ${^1}{\rm
S}_0$--state [10].

The first term in the effective coupling constant $G_{\rm \pi NN}$ comes
from the one--pion exchange for the squared momenta transfer $- q^2$ much
less than the squared pion mass $- q^2 \ll M^2_{\pi}$ and the subsequent
Fierz transformation of the nucleon fields [1,2]. We should emphasize that
due to Fierz transformation the effective NN interaction  caused by the
one--pion exchange contains a few contributions with different spinorial
structure, we have takeninto account only those terms which contribute to
the ${^1}{\rm S}_0$--state of the NN system. The second term is a
phenomenological one representing a collective contribution caused by the
integration over heavier meson fields like scalar mesons $\sigma(700)$,
$a_0(980)$ and $f_0(980)$, vector mesons $\rho(770)$ and $\omega(780)$ and
so on. This term is taken in the form used in the Effective Field Theory
(EFT) approach [11--13]. The effective interaction Eq.~(\ref{label1.1}) is
written in the isotopically invariant form, and the coupling constant
$G_{\rm \pi NN}$ can be never equal zero at $a_{\rm np} \neq 0$ due to
negative value of $a_{\rm np}$ imposed by nuclear forces, i.e., $a_{\rm np}
< 0$ [14].

In the low--energy limit the effective local four--nucleon interaction
Eq.~(\ref{label1.1}) vanishes due to the reduction
\begin{eqnarray}\label{label1.3}
[\bar{N}(x)\gamma_{\mu}\gamma^5 N^c (x)][\bar{N^c}(x)\gamma^{\mu}\gamma^5
N(x)] \to - [\bar{N}(x) \gamma^5 N^c(x)][\bar{N^c}(x) \gamma^5 N(x)],
\end{eqnarray}
where $N(x)$ is the neutron or the proton interpolating field. Such a
vanishing of the one--pion exchange contribution to the NN potential is
well--known in the EFT approach [11--13] and the PMA [14]. In power
counting [11--13] the interaction induced by the one--pion exchange is of
order $O(k^2)$, where $k$ is a relative momentum of the NN system. The
former is due the Dirac matrix $\gamma^5$ which leads to the  interaction
between small components of the Dirac bispinors of the nucleon wave
functions.

Thus, if either in the PMA or the EFT approach the effective local
four--nucleon interaction Eq.~(\ref{label1.1}) would be applied to the
description of the deuteron coupled to the nucleons, the contribution
would be scarcely significant. Therefore, both in the PMA and the EFT for
the correct description of strong low--energy nuclear forces one needs to
include an effective phenomenological NN potential, for instance, the
Argonne $v_{18}$ [15]. The one--pion exchange contribution is considered as
a perturbation.

In the RFMD due to the one--nucleon loop exchange the contributions of the
interactions $[\bar{N}(x)\gamma_{\mu}\gamma^5 N^c
(x)][\bar{N^c}(x)\gamma^{\mu}\gamma^5 N(x)]$ and $[\bar{N}(x) \gamma^5
N^c(x)][\bar{N^c}(x) \gamma^5 N(x)]$, or shortly $\gamma_{\mu}\gamma^5
\otimes \gamma^{\mu}\gamma^5$ and $\gamma^5 \otimes \gamma^5$, to the
amplitudes of nuclear  processes are different and do not cancel each other
in the low--energy limit. This is completely a peculiarity of one--nucleon
loop diagrams related to one--fermion loop anomalies [16,17,2]. For
instance, in the case of the neutron--proton radiative capture and the
photomagnetic disintegration of the deuteron the amplitudes of the
processes are defined by the triangle one--nucleon loop diagrams with  AVV
(axial--vector--vector) and PVV (pseudoscalar--vector--vector) vertices [2]
caused by $\gamma_{\mu}\gamma^5 \otimes \gamma^{\mu}\gamma^5$ and $\gamma^5
\otimes \gamma^5$ interactions, respectively. These diagrams are
well--known in particle physics in connection with  the Adler--Bell--Jackiw
axial anomaly [16] which plays a dominant role for the processes of the
decays $\pi^0 \to \gamma \gamma$, $\omega \to \pi^0 \gamma$ and so on [16].
The results of the calculation of these diagrams differ each other. Hence,
they give different contributions to the amplitudes of the processes and do
not cancel themselves in the low--energy limit.

Then, the amplitudes of the solar proton burning and the anti--neutrino
disintegration of the deuteron are defined by the one--nucleon loop
diagrams with AAV and APV vertices caused by $\gamma_{\mu}\gamma^5 \otimes
\gamma^{\mu}\gamma^5$ and $\gamma^5 \otimes \gamma^5$ interactions,
respectively. The contribution of the diagrams with APV vertices turns out
to be divergent and, therefore, negligibly small compared with the
contribution of the diagrams with AAV vertices [2], which contains
non--trivial convergent part related to one--fermion loop anomalies [2,17].

As a result in the low--energy limit amplitudes of nuclear processes
described by the RFMD contain only large components of Dirac bispinors of
wave functions of nucleons. This provides an effective enhancement of the
one--pion exchange for the description of strong low--energy interactions
of the nucleons in the ${^1}{\rm S}_0$--state. Thus, in the RFMD the
dynamics of strong low--energy nuclear interactions caused by one--nucleon
loop exchanges is the point of a dominant role  of the one--pion exchange
contribution to the effective low--energy interactions of nucleons in the
${^1}{\rm S}_0$--state.

The paper is organized as follows. In Sect.~2 we calculate the
astrophysical factor for the solar proton burning. In Sect.~3  we calculate
the cross section for the photomagnetic disintegration of the deuteron near
threshold and formulate the extrapolation procedure for the cross section
for energies far from threshold. In Sect.~4 we compute the cross section
for the anti--neutrino disintegration
of the deuteron  near threshold, extrapolate this cross section for
energies far from threshold and average the extrapolated cross section over
the reactor anti--neutrino energy spectrum.  In Conclusion we discuss the
obtained results and a justification of the RFMD. In Appendix A we adduce
the detailed calculation of the amplitude of the solar proton burning. In
Appendix B we show that in the RFMD one can describe low--energy elastic NN
scattering with non--zero effective range in accordance with low--energy
nuclear phenomenology.

\section{Astrophysical factor for the solar proton burning}
\setcounter{equation}{0}

In the RFMD the amplitude of the solar proton burning is defined by
one--nucleon loop diagrams [2]. The detailed calculation of the amplitude
of the solar proton burning is given in Appendix A. The result reads
\begin{eqnarray}\label{label2.1}
&&i{\cal M}({\rm p} + {\rm p} \to {\rm D} + {\rm e}^+ + \nu_{e}) = C(\eta)
\,G_{\rm V}\,g_{\rm A} M_{\rm N}\,G_{\rm \pi NN}\,\frac{3g_{\rm V}}{4\pi^2}\,
\nonumber\\
&&\hspace{1in}\times\,e^*_{\mu}(k_{\rm D})\,[\bar{u}(k_{\nu_{\rm
e}})\gamma^{\mu} (1-\gamma^5) v(k_{\rm e^+})]\,[\bar{u^c}(p_2) \gamma^5
u(p_1)],
\end{eqnarray}
where $G_{\rm V} = G_{\rm F}\cos \vartheta_{\rm C}$ with $G_{\rm
F}=1.166\times\,10^{-5}\,{\rm GeV}^{-2}$ and $\vartheta_{\rm C}$ are is the
Fermi weak coupling constant and the Cabibbo angle $\cos \vartheta_{\rm C}
= 0.975$. Then $g_{\rm A}=1.260\pm 0.012$ describes the renormalization of
the weak axial  hadron current by strong interactions [10], $g_{\rm V}$ is
the phenomenological coupling constant of the RFMD related to the electric
quadrupole moment of the deuteron: $g^2_{\rm V} = 2\pi^2 Q_{\rm D} M^2_{\rm
N}$ [2] with $Q_{\rm D} = 0.286\,{\rm fm}^2$ [10], $e^*_{\mu}(k_{\rm D})$
is a 4--vector of a polarization of the deuteron and $\bar{u}(k_{\nu_{\rm
e}})$, $v(k_{\rm e^+})$, $\bar{u^c}(p_2)$ and $u(p_1)$ are the Dirac
bispinors of neutrino, positron, and two protons, respectively. For the
binding energy of the deuteron we use the value $\varepsilon_{\rm D} =
2.225 \,{\rm MeV}$ [10]. The Coulomb repulsion between protons is taken
into account only in terms of the Gamow penetration factor [2,3] $C(\eta) =
\textstyle \sqrt{2\pi\eta}\,\exp(-\pi\eta)$ depending on the relative
velocity of the protons $v$ as $\eta =\alpha/v$ and $\alpha=1/137$ is the
fine structure constant.

The cross section for the low--energy p + p $\to$ D + ${\rm e}^+$ +
$\nu_{\rm e}$ reaction is defined
\begin{eqnarray}\label{label2.2}
\hspace{-0.5in}&&\sigma({\rm pp} \to {\rm D e^+ \nu_{\rm e}}) =
\frac{1}{v}\,\frac{1}{4E_1E_2}\int\,\overline{|{\cal M}({\rm p} + {\rm p}
\to {\rm D} + {\rm e}^+ + \nu_{\rm e})|^2}\,\nonumber\\
\hspace{-0.5in}&&\times (2\pi)^4\,\delta^{(4)}(k_{\rm D} + k_{\rm e^+ } +
k_{\nu_{\rm e}} - p_1 - p_2)\,\frac{d^3k_{\rm D}}{(2\pi)^3 2E_{\rm
D}}\frac{d^3k_{\rm e^+}}{(2\pi)^3 2E_{\rm e^+}}\frac{d^3k_{\nu_{\rm
e}}}{(2\pi)^3 2E_{\nu_{\rm e}}}\,,
\end{eqnarray}
where $v$ is a relative velocity of the protons and $E_i\,(i=1,2)$ are the
energies of the protons in the center of mass frame.

Then, $\overline{|{\cal M}({\rm p} + {\rm p} \to {\rm D} + {\rm e}^+
+\nu_{e})|^2}$ is the squared amplitude averaged over polarizations of
protons and summed over polarizations of final particles:
\begin{eqnarray}\label{label2.3}
\hspace{-0.5in}&&\overline{|{\cal M}({\rm p} + {\rm p} \to {\rm D} + {\rm
e}^+ + \nu_{e})|^2}= C^2(\eta)\,G^2_{\rm V} g^2_{\rm A} M^4_{\rm
N}\,G^2_{\rm \pi NN}\,\frac{9Q_{\rm D}}{8\pi^2} \times \Bigg(-
g^{\alpha\beta}+\frac{k^{\alpha}_{\rm D}k^{\beta}_{\rm D}}{M^2_{\rm
D}}\Bigg)\nonumber\\
\hspace{-0.5in}&&\times {\rm tr}\{(- m_{\rm e} + \hat{k}_{\rm
e^+})\gamma_{\alpha}(1-\gamma^5) \hat{k}_{\nu_{\rm
e}}\gamma_{\beta}(1-\gamma^5)\}\times \frac{1}{4}\times {\rm tr}\{(M_{\rm
N} - \hat{p}_2) \gamma^5 (M_{\rm N} + \hat{p}_1) \gamma^5\},
\end{eqnarray}
where $m_{\rm e}=0.511\;{\rm MeV}$ is the mass of positron, and we have
used the relation $g^2_{\rm V}/\pi^2 = 2\,Q_{\rm D}\,M^2_{\rm N}$ [2].

In the low--energy limit the computation of the traces yields
\begin{eqnarray}\label{label2.4}
\hspace{-0.5in}&&\Bigg(- g^{\alpha\beta}+\frac{k^{\alpha}_{\rm
D}k^{\beta}_{\rm D}}{M^2_{\rm D}}\Bigg)\,\times\,{\rm tr}\{( - m_{\rm e} +
\hat{k}_{\rm e^+})\gamma_{\alpha}(1-\gamma^5) \hat{k}_{\nu_{\rm
e}}\gamma_{\beta}(1-\gamma^5)\}= \nonumber\\
\hspace{-0.5in}&& = 24\,\Bigg( E_{\rm e^+} E_{\nu_{\rm e}} -
\frac{1}{3}\vec{k}_{\rm e^+}\cdot \vec{k}_{\nu_{\rm e}}\,\Bigg) ,\nonumber\\
\hspace{-0.5in}&&\frac{1}{4}\,\times\,{\rm tr}\{(M_{\rm N} - \hat{p}_2)
\gamma^5 (M_{\rm N} + \hat{p}_1) \gamma^5\} = 2\,M^2_{\rm N},
\end{eqnarray}
where we have neglected the relative kinetic energy of the protons with
respect to the mass of the proton.

Substituting Eq.~(\ref{label2.4}) in Eq.~(\ref{label2.3}) we get
\begin{eqnarray}\label{label2.5}
\overline{|{\cal M}({\rm p} + {\rm p} \to {\rm D} + {\rm e}^+ +
\nu_{e})|^2} = C^2(\eta)\,G^2_{\rm V}\, g^2_{\rm A} M^6_{\rm N}\,G^2_{\rm
\pi NN}\,\frac{54 Q_{\rm D}}{\pi^2}\Bigg( E_{\rm e^+} E_{\nu_{\rm e}} -
\frac{1}{3}\vec{k}_{\rm e^+}\cdot \vec{k}_{\nu_{\rm e}}\Bigg).
\end{eqnarray}
The integration over the phase volume of the final ${\rm D}{\rm e}^+
\nu_{\rm e}$--state we perform in the non--relativistic limit
\begin{eqnarray}\label{label2.6}
\hspace{-0.5in}&&\int\frac{d^3k_{\rm D}}{(2\pi)^3 2E_{\rm
D}}\frac{d^3k_{\rm e^+}}{(2\pi)^3 2E_{\rm e^+}}\frac{d^3k_{\nu_{\rm
e}}}{(2\pi)^3 2 E_{\nu_{\rm e}}}\,(2\pi)^4\,\delta^{(4)}(k_{\rm D} +
k_{\ell} - p_1 - p_2)\,\Bigg( E_{\rm e^+} E_{\nu_{\rm e}} -
\frac{1}{3}\vec{k}_{\rm e^+}\cdot \vec{k}_{\nu_{\rm e}}\,\Bigg)\nonumber\\
\hspace{-0.5in}&&= \frac{1}{32\pi^3 M_{\rm N}}\,\int^{W + T_{\rm
pp}}_{m_{\rm e}}\sqrt{E^2_{\rm e^+}-m^2_{\rm e}}\,E_{\rm e^+}(W + T_{\rm
pp} - E_{\rm e^+})^2\,d E_{\rm e^+} = \frac{(W + T_{\rm pp})^5}{960\pi^3
M_{\rm N}}\,f(\xi),
\end{eqnarray}
where $W = \varepsilon_{\rm D} - (M_{\rm n} - M_{\rm p}) = (2.225
-1.293)\,{\rm MeV} = 0.932\,{\rm MeV}$, $T_{\rm pp} = M_{\rm N}\,v^2/4$ is
the kinetic energy of the relative movement of the protons, and  $\xi =
m_{\rm e}/(W + T_{\rm pp})$. The function $f(\xi)$ is defined by the
integral
\begin{eqnarray}\label{label2.7}
\hspace{-0.5in}f(\xi)&=&30\,\int^1_{\xi}\sqrt{x^2 -\xi^2}\,x\,(1-x)^2 dx=(1
- \frac{9}{2}\,\xi^2 - 4\,\xi^4)\,\sqrt{1-\xi^2}\nonumber\\
\hspace{-0.5in}&&+ \frac{15}{2}\,\xi^4\,{\ell
n}\Bigg(\frac{1+\sqrt{1-\xi^2}}{\xi}\Bigg)\Bigg|_{T_{\rm pp} = 0} =  0.222
\end{eqnarray}
and normalized to unity at $\xi = 0$.

Thus, the cross section for the solar proton burning is given by
\begin{eqnarray}\label{label2.8}
\sigma({\rm pp} \to {\rm D e^+\nu_{\rm e}}) &=& \frac{e^{\displaystyle -
2\pi\eta}}{v^2}\,\alpha\,\frac{9g^2_{\rm A} G^2_{\rm V} Q_{\rm D} M^3_{\rm
N}}{320\,\pi^4}\,G^2_{\rm \pi NN}\,(W + T_{\rm pp})^5\,f\Bigg(\frac{m_{\rm
e}}{W + T_{\rm pp}}\Bigg)=\nonumber\\
&=& \frac{S_{\rm pp}(T_{\rm pp})}{T_{\rm pp}}\,e^{\displaystyle - 2\pi\eta}.
\end{eqnarray}
The astrophysical factor $S_{\rm pp}(T_{\rm pp})$ reads
\begin{eqnarray}\label{label2.9}
S_{\rm pp}(T_{\rm pp}) = \alpha\,\frac{9g^2_{\rm A}G^2_{\rm V}Q_{\rm
D}M^4_{\rm N}}
{1280\pi^4}\,G^2_{\rm \pi NN}\,(W + T_{\rm pp})^5\,f\Bigg(\frac{m_{\rm
e}}{W + T_{\rm pp}}
\Bigg),
\end{eqnarray}
At zero kinetic energy of the protons $T_{\rm pp} = 0$ the astrophysical
factor $S_{\rm pp}(0)$ is given by
\begin{eqnarray}\label{label2.10}
\hspace{-0.5in}S_{\rm pp}(0) =\alpha\,\frac{9g^2_{\rm A}G^2_{\rm V}Q_{\rm
D}M^4_{\rm N}}{1280\pi^4}\,G^2_{\rm \pi NN}\,\,W^5\,f\Bigg(\frac{m_{\rm
e}}{W}\Bigg) =  4.02\,\times 10^{-25}\,{\rm MeV\,\rm b}.
\end{eqnarray}
The value $S_{\rm pp}(0) = 4.02 \times 10^{-25}\,{\rm MeV\,\rm b}$ agrees
with the value $S_{\rm pp}(0) = 3.89 \times 10^{-25}\,{\rm MeV\,\rm b}$
obtained by Kamionkowski and Bahcall in the PMA [18] and $S_{\rm pp}(0) =
4.05 \times 10^{-25}\,{\rm MeV\,\rm b}$ having been calculated recently in
the EFT approach [19].

Since due to charge independence of the weak interaction strength the
reaction of the anti--neutrino disintegration of the deuteron
$\bar{\nu}_{\rm e}$ + D $\to$ e$^+$ + n + n is valued  as a terrestrial
equivalent of the solar proton burning, for the justification of the
validity of our result Eq.~(\ref{label2.10}) we suggest to calculate the
cross section for the disintegration of the deuteron by reactor
anti--neutrinos $\bar{\nu}_{\rm e}$ + D $\to$ e$^+$ + n + n. This
calculation is carried out in Sect.~4.  However, the RFMD with the
effective local four--nucleon interaction Eq.~(\ref{label1.1}) allows to
calculate the cross section for the reaction $\bar{\nu}_{\rm e}$ + D $\to$
e$^+$ + n + n  for  anti--neutrino energies close to threshold. As it is
shown in Sect.~4 this cross section agrees well with the result obtained in
the PMA.  For the comparison of our result with the experimental data
represented in the form of the cross section averaged over the
anti--neutrino energy spectrum we need to extrapolate the cross section
calculated near threshold for the anti--neutrino energies far from
threshold. In order to formulate such an extrapolation procedure we turn to
the consideration of the photomagnetic disintegration of the deuteron
$\gamma$ + D $\to$ n + p.

\section{Photomagnetic disintegration of the deuteron}
\setcounter{equation}{0}

First, let us consider the cross section for the photomagnetic
disintegration of the deuteron $\gamma$ + D $\to$ n + p near threshold. The
cross section can be easily obtained by using the results of Ref.~[2,3]. It
reads
\begin{eqnarray}\label{label3.1}
\sigma^{\rm \gamma D}(\omega) = \sigma_0
\,\Bigg(\frac{\omega}{\varepsilon_{\rm D}}\Bigg)\,
k r_{\rm D},
\end{eqnarray}
where $k=\sqrt{M_{\rm N}(\omega - \varepsilon_{\rm D})}$ is the relative
momentum of the np system, $\omega$ is the energy of the photon and $r_{\rm
D} = 1/\sqrt{\varepsilon_{\rm D}M_{\rm N}} = 4.315\;{\rm fm}$ is the radius
of the deuteron, and $\sigma_0$ is given by
\begin{eqnarray}\label{label3.2}
\hspace{-0.5in}\sigma_0 = (\mu_{\rm p}-\mu_{\rm n})^2\frac{25\alpha Q_{\rm
D}}{192\pi^2}
G^2_{\rm \pi NN}\,\varepsilon^{3/2}_{\rm D}M^{5/2}_{\rm N} = 5.9\times
10^{-27}\,{\rm cm}^2.
\end{eqnarray}
The cross section $\sigma^{\rm \gamma D}(\omega)$, calculated in the PMA
near threshold has the same form as Eq.~(\ref{label3.1}) but with
$\sigma_0$ given by [14]
\begin{eqnarray}\label{label3.3}
\hspace{-0.5in} \sigma_0 = \frac{2\pi\alpha}{3 M^2_{\rm N}}\,(\mu_{\rm
p}-\mu_{\rm n})^2
\Big(1 - a_{\rm np}\sqrt
{\varepsilon_{\rm D} M_{\rm N}}\Big)^2 = 6.3\times 10^{-27}\,{\rm cm}^2.
\end{eqnarray}
It is seen that $\sigma_0$  defined by Eq.~(\ref{label3.2}) and
Eq.~(\ref{label3.3}) agree within an accuracy better than 10$\%$. The
analogous agreement can be drawn out from the comparison of the
reaction rates for the neutron--proton radiative capture for thermal
neutrons [3,14]:
\begin{eqnarray}\label{label3.4}
\hspace{-0.5in}v\sigma^{\rm np}(k) = \displaystyle
\left\{\begin{array}{r}\displaystyle
(\mu_{\rm p}-\mu_{\rm n})^2
\frac{25\alpha Q_{\rm D}}{64\pi^2}\,G^2_{\rm \pi NN}\,
\varepsilon^3_{\rm D}M_{\rm N} = 2.0\times 10^{-30}{\rm cm}^2,\\
\displaystyle (\mu_{\rm p}-\mu_{\rm n})^2 \frac{2\pi \alpha}{M^2_{\rm N}}
\Big(1 - a_{\rm np}
\sqrt{\varepsilon_{\rm D} M_{\rm N}}\Big)^{\!2}
\Bigg(\frac{\varepsilon_{\rm D}}{M_{\rm N}}
\Bigg)^{\!3/2} = 2.2\times 10^{-30}{\rm cm}^2
,\end{array}\right.
\end{eqnarray}
where $v$ and $k$ are a relative velocity and a relative 3--momentum of the
np system. Thus, near threshold the RFMD supplemented by the effective
local four--nucleon interaction Eq.~(\ref{label1.1}) provides a dynamics of
strong low--energy nuclear forces describing the cross sections for the
neutron--proton radiative capture and the photomagnetic disintegration of
the deuteron in good agreement with the PMA.

Now let us proceed to the formulation of the extrapolation procedure. For
this aim we suggest to consider the cross section for the photomagnetic
disintegration of the deuteron calculated in the PMA for the photon
energies far from threshold [14]
\begin{eqnarray}\label{label3.5}
\sigma^{\rm \gamma D}(\omega) = \sigma_0
\Bigg(\frac{\omega}{\varepsilon_{\rm D}}\Bigg)
\frac{k r_{\rm D}}{1 +  r^2_{\rm D}k^2}\frac{1}{1 + a^2_{\rm np} k^2} =
\sigma_0
\Bigg(\frac{\omega}{\varepsilon_{\rm D}}\Bigg)\,k r_{\rm D}F^{\rm D}_{\rm
np}(k^2),
\end{eqnarray}
where the function $F^{\rm D}_{\rm np}(k^2)$ is defined as
\begin{eqnarray}\label{label3.6}
F^{\rm D}_{\rm np}(k^2) = \frac{1}{1 +  r^2_{\rm D}k^2}\frac{1}{1 +
a^2_{\rm np} k^2}.
\end{eqnarray}
The cross section Eq.~(\ref{label3.5}) differs from the cross section
Eq.~(\ref{label3.1}) calculated near threshold by the factor $F^{\rm
D}_{\rm np}(k^2)$, which can be considered as a form factor taking into
account a spatial smearing of the deuteron through the factor $1/(1 +
r^2_{\rm D}k^2)$ and  the np system through the factor $1/(1 +  a^2_{\rm
np}k^2)$.

In order not to come into the contradiction with the PMA we suggest to
extrapolate the cross section Eq.~(\ref{label3.1}) for the photon energies
far from threshold by means of the form factor $F^{\rm D}_{\rm np}(k^2)$.
This extrapolation assumes that the cross section for the photomagnetic
disintegration of the deuteron calculated in the RFMD has the form
Eq.~(\ref{label3.5}) with $\sigma_0$ defined by Eq.~(\ref{label3.2}). Such
an extrapolation applied to the cross section for the anti--neutrino
disintegration of the deuteron should lead to the appearance of the form
factor
\begin{eqnarray}\label{label3.7}
F^{\rm D}_{\rm nn}(k^2) = \frac{1}{1 +  r^2_{\rm D}k^2}\frac{1}{1 +
a^2_{\rm nn} k^2}
\end{eqnarray}
describing a spatial smearing of the deuteron through the factor $1/(1 +
r^2_{\rm D}k^2)$ and the
nn system through the factor $1/(1 +  a^2_{\rm nn}k^2)$, where $a_{\rm nn}
= - 17\,{\rm fm}$ is the S--wave scattering length of the nn scattering in
the ${^1}{\rm S}_0$--state [10].

We should accentuate that we have discussed the photomagnetic
disintegration of the deuteron in order to draw out the hint for the
formulation of the extrapolation procedure for the cross section for the
disintegration of the deuteron by anti--neutrinos. We are not aiming here
to compute a total cross section for the disintegration of the deuteron by
photons. In fact, it is well known [14] that a photomagnetic part
predominates only at small relative momenta of the np pair at $k\,r_{\rm
D}\ll 1$ and the photoelectric part becomes important at $k\,r_{\rm D}\ge
1$, i.e., $\omega \ge  \varepsilon_{\rm D} = 4.45\,{\rm MeV}$.

Also the computation of the reaction rate for the neutron--proton radiative
capture given by Eq.~(\ref{label3.4}) can be considered as a lowest
approximation. Certainly, the cross section $\sigma^{\rm np}(k) = 276\,{\rm
mb}$ computed for thermal neutrons at laboratory velocities $v/c =
7.34\cdot 10^{\,-\,6}$ (the absolute value is $v=2.2\cdot 10^{\,5}\,{\rm
cm/sec}$) [3] agrees reasonably well with the experimental data
$\sigma^{\rm np}_{\exp}(k) = (334.2 \pm 0.5)\,{\rm mb}$ [20]. However, the
central theoretical value is 20$\%$ smaller than the central experimental
value. Such a problem has been solved  for the first time within the PMA by
Riska and Brown [21] who showed that the discrepancy of order 10$\%$
between the theoretical value of the cross section $\sigma^{\rm np}(k) = (302.5\pm 4.0)\,{\rm mb}$ [14] calculated in the PMA and the experimental value $
\sigma^{\rm np}_{\exp}(k)\,=\,(334.2 \pm 0.5)\,{\rm mb}$ can be explained
by exchange--current contributions. In the EFT approach and Chiral
perturbation theory the same result has been obtained by Park {\it et
al.\,} [22]. The investigation of such fine effects of the neutron--proton
radiative capture and incorporation of Chiral perturbation theory into the
RFMD is in the programme of further applications of the RFMD to the
processes of low--energy interactions of the deuteron.

\section{Anti--neutrino disintegration of the deuteron}
\setcounter{equation}{0}

The calculation of the amplitude of the anti--neutrino disintegration of
the deuteron is analogous to the calculation of the amplitude of the solar
proton burning. The effective Lagrangian responsible for the transition
$\bar{\nu}_{\rm e}$ + D $\to$ e$^+$ + n + n is given by Eq.~({\rm A}.49)
$$
{\cal L}_{\rm \bar{\nu}_{\rm e}D \to e^+ nn }(x) = g_{\rm A}G_{\rm \pi
NN}\frac{G_{\rm V}}{\sqrt{2}}\frac{3g_{\rm
V}}{8\pi^2}\,D_{\mu\nu}(x)\,[\bar{n}(x)\gamma^{\mu}\gamma^5
n^c(x)]\,[\bar{\psi}_{\nu_{\rm e}}(x)\gamma^{\nu}(1 - \gamma^5) \psi_{\rm
e}(x)].
$$
The amplitude of the $\bar{\nu}_{\rm e}$ + D $\to$ e$^+$ + n + n process reads
\begin{eqnarray}\label{label4.1}
\hspace{-0.5in}i{\cal M}(\bar{\nu}_{\rm e} + {\rm D} \to {\rm
e}^+ + {\rm n} + {\rm n}) &=& -\,g_{\rm A} M_{\rm N}\,G_{\rm \pi
NN}\,\frac{G_{\rm
V}}{\sqrt{2}}\,\frac{3g_{\rm V}}{2\pi^2}\,\nonumber\\
&&\times e_{\mu}(Q)\,[\bar{v}(k_{\bar{\nu}_{\rm
e}})\gamma^{\mu}(1-\gamma^5)
v(k_{{\rm e}^+})]\,[\bar{u}(p_1) \gamma^5 u^c(p_2)],
\end{eqnarray}
where $\bar{v}(k_{\bar{\nu}_{\rm e}})$, $v(k_{{\rm e}^+})$, $\bar{u}(p_1)$
and $u^c(p_2)$ are the Dirac bispinors of the anti--neutrino, positron and
neutrons, $e_{\mu}(Q)$ is the 4--vector
of the polarization of the deuteron. We have taken into account that
$\bar{u}(p_2) \gamma^5 u^c(p_1) = -\bar{u}(p_1) \gamma^5 u^c(p_2)$. The
amplitude Eq.~(\ref{label4.1}) squared, averaged over polarizations of the
deuteron and summed over polarizations of the
final particles reads
\begin{eqnarray}\label{label4.2}
\overline{|{\cal M}(\bar{\nu}_{\rm e} + {\rm D} \to  {\rm
e}^+ + {\rm n} + {\rm n})|^2} = g^2_{\rm A}M^6_{\rm N}G^2_{\rm \pi NN}\,
\frac{144 G^2_{\rm V}Q_{\rm D}}{\pi^2}\Bigg( E_{{\rm e}^+}
E_{\bar{\nu}_{\rm e}} - \frac{1}{3}\vec{k}_{{\rm e}^+}\cdot
\vec{k}_{\bar{\nu}_{\rm e}}\Bigg).
\end{eqnarray}
Due to charge independence of the weak interaction strength the matrix
element Eq.~(\ref{label4.2}) is related to the matrix element of the solar
proton burning Eq.~(\ref{label2.5}) by the relation
\begin{eqnarray}\label{label4.3}
\overline{|{\cal M}(\bar{\nu}_{\rm e} + {\rm D} \to {\rm
e}^+ + {\rm n} + {\rm n})|^2} = \frac{8}{3} \times \frac{1}{\displaystyle
C^2(\eta)} \times \overline{|{\cal M}({\rm p} + {\rm p} \to {\rm D}  + {\rm
e}^+ + \nu_{\rm e})|^2},
\end{eqnarray}
where $8/3$ is a combinatorial factor. This relation means that the
dynamics of strong low--energy nuclear forces governing the processes of
the anti--neutrino disintegration of the deuteron $\bar{\nu}_{\rm e}$ + D
$\to$ n + n + e$^+$ and the solar proton burning p + p $\to$ D + e$^+$ +
$\nu_{\rm e}$ should have the same origin.

The expression Eq.~(\ref{label4.3}) extrapolated for energies far from
threshold according to the procedure suggested in Sect.~3 reads
\begin{eqnarray}\label{label4.4}
\overline{|{\cal M}(\bar{\nu}_{\rm e} + {\rm D} \to  {\rm e}^+ + {\rm n} +
{\rm n})|^2} =
g^2_{\rm A}M^6_{\rm N}G^2_{\rm \pi NN}\frac{144 G^2_{\rm V}Q_{\rm
D}}{\pi^2}\Bigg( E_{{\rm e}^+} E_{\bar{\nu}_{\rm e}} -
\frac{1}{3}\vec{k}_{{\rm e}^+} \cdot \vec{k}_{\bar{\nu}_{\rm e}}\Bigg)
 F^{\rm D}_{\rm nn}(k^2).
\end{eqnarray}
The form factor $F^{\rm D}_{\rm nn}(k^2)$, describing a spatial smearing of
the deuteron and the nn system, is given by Eq.~(\ref{label3.7})
\begin{eqnarray*}
F^{\rm D}_{\rm nn}(k^2) = \frac{1}{1 + r^2_{\rm D}k^2}\,\frac{1}{1 +
a^2_{\rm nn}\,k^2},
\end{eqnarray*}
where $k = \sqrt{M_{\rm N}T_{\rm nn}}\,$ is the relative momentum and
$T_{\rm nn}$ is the kinetic energy of the relative movement of the nn
system, and $a_{\rm nn} = -17\,{\rm fm}$ [10] is the S--wave scattering
length of the low--energy elastic nn scattering in the ${^1}{\rm
S}_0$--state. A much more complicated extrapolation of the form factors of
the amplitude of the disintegration of the deuteron by anti--neutrinos has
been suggested by Mintz [23].

The cross section for the process $\bar{\nu}_{\rm e}$ + D $\to$ e$^+$ + n +
n  is defined by
\begin{eqnarray}\label{label4.5}
&&\sigma^{\bar{\nu}_{\rm e} D}(E_{\bar{\nu}_{\rm e}}) =
\frac{1}{4E_{\rm D}E_{\bar{\nu}_{\rm e}}}\int\,\overline{|{\cal
M}(\bar{\nu}_{\rm e} +
{\rm D} \to  {\rm e}^+ + {\rm n} + {\rm n})|^2}\nonumber\\
&&\frac{1}{2}\,(2\pi)^4\,\delta^{(4)}(Q + k_{{\bar{\nu}_{\rm e}}} - p_1 -
p_2 - k_{{\rm e}^+})\,
\frac{d^3p_1}{(2\pi)^3 2E_1}\frac{d^3 p_2}{(2\pi)^3 2E_2}\frac{d^3k_{{\rm
e}^+}}{(2\pi)^3
2E_{{\rm e}^+}},
\end{eqnarray}
where $E_{\rm D}$, $E_{\bar{\nu}_{\rm e}}$, $E_1$, $E_2$  and $E_{{\rm
e}^+}$ are the energies of the deuteron, the anti--neutrino, the neutrons
and the positron. The integration over the phase volume of
the (${\rm n n e^+}$)--state we perform in the non--relativistic limit and
in the rest frame of the deuteron,
\begin{eqnarray}\label{label4.6}
&&\frac{1}{2}\,\int\frac{d^3p_1}{(2\pi)^3 2E_1}\frac{d^3p_2}{(2\pi)^3 2E_2}
\frac{d^3k_{{\rm e}^+}}{(2\pi)^3 2E_{{\rm e}^+}}(2\pi)^4\,\delta^{(4)}(Q +
k_{{\bar{\nu}_{\rm e}}} - p_1 - p_2 - k_{{\rm e}^+})\,\nonumber\\
&&\hspace{1in} \times\, \Bigg( E_{{\rm e}^+} E_{\bar{\nu}_{\rm e}} -
\frac{1}{3}
\vec{k}_{{\rm e}^+}\cdot
\vec{k}_{\bar{\nu}_{\rm e}}\Bigg)\,F^{\rm D}_{\rm nn}(M_{\rm N}\,T_{\rm
nn}) =\nonumber\\
&&= \frac{E_{\bar{\nu}_{\rm e}}M^3_{\rm N}}{1024\pi^2}\,\Bigg(\frac{E_{\rm
th}}{M_{\rm N}}
\Bigg)^{\!\!7/2}\Bigg(\frac{2 m_{\rm e}}{E_{\rm
th}}\Bigg)^{\!\!3/2}\frac{8}{\pi E^2_{\rm th}}
\int\!\!\!\int dT_{\rm e^+} dT_{\rm nn}\,
\sqrt{T_{\rm e^+}T_{\rm nn}}\, \nonumber\\
&&\times\,F^{\rm D}_{\rm nn}(M_{\rm N}\,T_{\rm nn})\,\Bigg(1 + \frac{T_{\rm
e^+}}{m_{\rm e}}
\Bigg)\,{\displaystyle \sqrt{1 + \frac{T_{\rm e^+}}{2 m_{\rm e}} }}\,
\delta\Big(E_{\bar{\nu}_{\rm e}}- E_{\rm th} - T_{\rm e^+}
 - T_{\rm nn}\Big) = \nonumber\\
&&= \frac{E_{\bar{\nu}_{\rm e}}M^3_{\rm N}}{1024\pi^2}\,\Bigg(\frac{E_{\rm
th}}{M_{\rm N}}
\Bigg)^{\!\!7/2}\Bigg(\frac{2 m_{\rm e}}{E_{\rm th}}\Bigg)^{\!\!3/2}\Bigg
(\frac{E_{\bar{\nu}_{\rm e}}}
{E_{\rm th}} - 1\Bigg)^{\!\!2}\,f\Bigg(\frac{E_{\bar{\nu}_{\rm e}}}
{E_{\rm th}}\Bigg),
\end{eqnarray}
where $T_{\rm e^+}$ and $m_{\rm e} = 0.511\,{\rm MeV}$ are the kinetic
energy and the mass of
the positron,
$E_{\rm th}$ is the anti--neutrino energy threshold of the reaction
$\bar{\nu}_{\rm e}$ + D $\to$ e$^+$ + n + n and is given by $E_{\rm th}=
\varepsilon_{\rm D} + m_{\rm e} + (M_{\rm n} - M_{\rm p}) =
(2.225 + 0.511 + 1.293) \, {\rm MeV} =
4.029\,{\rm MeV}$. The function $f(y)$, where $y=E_{\bar{\nu}_{\rm
e}}/E_{\rm th}$,
is defined as
\begin{eqnarray}\label{label4.7}
&&f(y) = \frac{8}{\pi}\,\int\limits^{1}_{0} dx\,\sqrt{x\,(1 - x)}\,F^{\rm
D}_{\rm nn}
(M_{\rm N}E_{\rm th}\,(y - 1)\,x)\nonumber\\
&&\times\,\Bigg(1 + \frac{E_{\rm th}}{m_{\rm e}}(y-1)(1-x)\Bigg) \,\sqrt{1
+ \frac{E_{\rm th}}
{2 m_{\rm e}}(y-1)(1-x)},
\end{eqnarray}
where we have changed the variable $T_{\rm nn} = (E_{\bar{\nu}_{\rm e}} -
E_{\rm th})\,x$.
The function $f(y)$ is normalized to unity at $y=1$, i.e., at threshold
$E_{\bar{\nu}_{\rm e}}= E_{\rm th}$. Thus, the cross section for the
anti--neutrino disintegration
of the deuteron reads
\begin{eqnarray}\label{label4.8}
\sigma^{\rm \bar{\nu}_{\rm e} D}(E_{\bar{\nu}_{\rm e}}) = \sigma_0\,(y -
1)^2\,f(y),
\end{eqnarray}
where $\sigma_0$ is defined by
\begin{eqnarray}\label{label4.9}
\hspace{-0.5in}\sigma_0 = Q_{\rm D}\,G^2_{\rm \pi NN}\,\frac{9g^2_{\rm A}
G^2_{\rm V} M^8_{\rm N}}{512\pi^4}\,\Bigg(\frac{E_{\rm th}}{M_{\rm
N}}\Bigg)^{\!\!7/2}\Bigg(\frac{2 m_{\rm e}}{E_{\rm th}}\Bigg)^{\!\!3/2}
=(4.53\pm 0.86) \times \,10^{-43}\,{\rm cm}^2.
\end{eqnarray}
Here $\pm 0.86$ describes the assumed theoretical uncertainty of the RFMD
which is about 19\,$\%$\footnote{This is the improved estimate of the
theoretical uncertainty. The former was 30$\%$ [3].}. The value $\sigma_0 =
(4.53 \pm 0.86)\times\,10^{-43} {\rm cm}^2$ agrees  with the value
$\sigma_0 = (4.68 \pm 1.14) \times \,10^{-43}\,{\rm cm}^2$ obtained in the
PMA [24,25] (see Fig.~7 of Ref.~[4]).

The experimental data on the anti--neutrino disintegration of the deuteron
are given in terms of the cross section averaged over the reactor
anti--neutrino energy spectrum  per anti--neutrino
fission in the energy region of anti--neutrinos $E_{\rm th} \le
E_{\bar{\nu}_{\rm e}} \le 10\,{\rm MeV}$:
$<\sigma^{\bar{\nu}_{\rm e}D}(E_{\bar{\nu}_{\rm e}})>_{\exp} = (1.5\pm 0.4)
\times 10^{-45}\,{\rm cm}^2/{\bar{\nu}_{\rm e}}\,{\rm fission}$ [7],
$<\sigma^{\bar{\nu}_{\rm e}D}(E_{\bar{\nu}_{\rm e}})>_{\exp} = (0.9 \pm
0.4) \times 10^{-45}\, {\rm cm}^2/
{\bar{\nu}_{\rm e}}\, {\rm fission}$ [8] and $<\sigma^{\bar{\nu}_{\rm
e}D}(E_{\bar{\nu}_{\rm e}})>_{\exp} = (1.84\pm 0.04) \times 10^{-45}\,{\rm
cm}^2/{\bar{\nu}_{\rm e}}\,{\rm fission}$ [9].

The cross section $<\sigma^{\rm \bar{\nu}_{\rm e} D}(E_{\bar{\nu}_{\rm
e}})>$, calculated in the RFMD, extrapolated and averaged over the reactor
anti--neutrino
Avignone--Greenwood energy spectrum [5,6] in the energy region $E_{\rm th}\le
E_{\bar{\nu}_{\rm e}} \le 10\;{\rm MeV}$, is given by
\begin{eqnarray}\label{label4.10}
<\sigma^{\rm \bar{\nu}_{\rm e} D}(E_{\bar{\nu}_{\rm e}})> &=&
\frac{a}{N_{\bar{\nu}_{\rm e}}}\int\limits^{2.482}_{1}dy\,e^{\displaystyle
-b\,y}\,
\sigma_0\,(y-1)^2\,f(y) = \nonumber\\
&=& (2.10 \pm 0.40)\times 10^{-45}\,{\rm cm}^2/\,\bar{\nu}_{\rm e}\,{\rm
fission},
\end{eqnarray}
where $a = 17.8\, E_{\rm th} = 71.72$, $b = 1.01\, E_{\rm th} = 4.07$, and
$N_{\bar{\nu}_{\rm e}} = 6$ is the number of anti--neutrinos per fission
[5,6]. The theoretical value Eq.~(\ref{label4.10}) agrees  well with the
experimental values given by Reines {\it et al.} [7] and Russian groups
[9], while the agreement with the value given by Reines {\it et al.}~[8] is
only qualitative.

Thus, the calculation of the cross section for the anti--neutrino
disintegration of the deuteron in agreement with the PMA and the
experimental data confirms our result obtained for the astrophysical factor
$S_{\rm pp}(0) = 4.02 \times 10^{-25}\,{\rm MeV\,b}$ for the solar proton
burning.

\section{Conclusion}
\setcounter{equation}{0}

We have shown that the RFMD supplemented by the local four--nucleon
interaction Eq.~(\ref{label1.1}) describes the processes of the solar
proton burning p + p $\to$ D + e$^+$ + $\nu_{\rm e}$, the neutron--proton
radiative capture n + p $\to$ D + $\gamma$, the disintegration of the
deuteron by photons $\gamma$ + D $\to$ n + p and  anti--neutrinos
$\bar{\nu}_{\rm e}$ + D $\to$ e$^+$ + n + n in good agreement with the PMA.
The astrophysical factor $S_{\rm pp}(0) = 4.02\times\,10^{-25}\,{\rm
MeV\,b}$ for the solar proton burning, the reaction rates for the
neutron--proton radiative capture and the cross sections for the
disintegration of the deuteron by photons and anti--neutrinos calculated
near thresholds of the reactions agree with the results obtained in the PMA
within an accuracy better than 10$\%$.

In order to compare our result for the cross section for the disintegration
of deuteron by anti--neutrinos with experimental data we suggested the
procedure of the extrapolation of the cross section calculated near
threshold to the energy region far from threshold. Of course, such an
extrapolation is not unique.  Therefore, we have suggested to formulate the
extrapolation procedure fitting the cross section for the photomagnetic
disintegration of the deuteron calculated in the PMA. This extrapolation
assumes the multiplication of the cross section, calculated in the RFMD near
threshold, by the form factor
\begin{eqnarray}\label{label5.1}
F^{\rm D}_{\rm NN}(k^2) = \frac{1}{1 +  r^2_{\rm D}k^2}\frac{1}{1 +
a^2_{\rm NN} k^2}
\end{eqnarray}
describing a spatial smearing of the deuteron by a factor $1/(1 +
r^2_{\rm D}k^2)$ and the NN system by a factor $1/(1 +  a^2_{\rm NN}k^2)$,
where $k$ is a relative momentum of the NN system. A much more complicated
extrapolation of the amplitude of the process $\bar{\nu}_{\rm e}$ + D $\to$
e$^+$ + n + n has been suggested by Mintz [23].

The extrapolated cross section for the disintegration of the deuteron by
antineutrinos averaged over the reactor anti--neutrino energy spectrum
$<\sigma^{\rm \bar{\nu}_{\rm e} D}(E_{\bar{\nu}_{\rm e}})>
= (2.10 \pm 0.40)\times 10^{-45}\,{\rm cm}^2/\,\bar{\nu}_{\rm e}\,{\rm
fission}$ agrees well with the experimental data [7,9].

The cross section for the disintegration of the deuteron by anti--neutrinos
$\bar{\nu}_{\rm e}$ + D $\to$ e$^+$ + n + n in dependence on the
anti--neutrino energy has been calculated recently in the PMA in Ref.~[26].
Since  near threshold our cross section agrees well with the PMA result,
for the verification of the extrapolation procedure we can, say, compare
the cross section at $E_{\bar{\nu}_{\rm e}} = 10\,{\rm MeV}$:
\begin{eqnarray}\label{label5.2}
\sigma^{\rm \bar{\nu}_{\rm e} D}(E_{\bar{\nu}_{\rm
e}})|_{\displaystyle E_{\bar{\nu}_{\rm e}}=10\,{\rm MeV}} &=& 1.02\times
10^{-42}\,{\rm cm}^2,\,{\rm RFMD},\nonumber\\
\sigma^{\rm \bar{\nu}_{\rm e} D}(E_{\bar{\nu}_{\rm e}})|_{\displaystyle
E_{\bar{\nu}_{\rm e}} = 10\,{\rm MeV}} &=& 1.13 \times 10^{-42}\,{\rm
cm}^2,\,{\rm PMA}.
\end{eqnarray}
The visible agreement with the PMA result can serve too as a confirmation
of a validity of our extrapolation procedure.

The RFMD describing a  relative movement of the nucleons inside the
deuteron in terms of one--nucleon loop exchanges suggests a dynamics of
strong low--energy nuclear forces completely different to the PMA and the
EFT approach. The neutron--proton--deuteron vertices are point--like and
defined by a phenomenological {\it local} conserving nucleon current
$J^{\mu}(x) = - ig_{\rm V}[\bar{p}(x) \gamma^\mu n^c(x) - \bar{n}(x)
\gamma^\mu p^c(x)]$ , i.e., $\partial_{\mu} J^{\mu}(x) = 0$, accounting for
spinorial and isotopical properties of the deuteron, and $g_{\rm V}$ is a
dimensionless phenomenological coupling constant. The deuteron  is
represented by a local field operator $D_{\mu}(x)$ (or
$D^{\dagger}_{\mu}(x)$), the action of which on a vacuum state annihilates
(or creates) the deuteron. The low--energy parameters of the deuteron such
as the binding energy $\varepsilon_{\rm D}$, the electric quadrupole
$Q_{\rm D}$ and the anomalous magnetic dipole $\kappa_{\rm D}$ moments are
induced by vacuum fluctuations of the neutron and the proton fields in a
quantum field theory way. The description of strong low--energy nuclear
forces in terms of one--nucleon loop exchanges provides opportunity to
convey in nuclear physics a huge experience of fermion anomalies [16,17,27]
which had been stored in particle physics from the paper by Adler [16]
concerning the derivation of the anomalous contribution to the axial  Ward
identity. In the area of low--energy interactions of low--lying mesons such
an experience has been focused mainly upon the derivation of anomalous
contributions [28--30] to Effective Chiral Lagrangians [31].

The main problem which we encounter for the practical realization of the
derivation of effective Lagrangians of low--energy interactions of the
deuteron through one--nucleon loop exchanges lies in the necessity to
satisfy requirement of {\it locality} of these interactions  related to
{\it the condition of microscopic causality} in a quantum field theory
approach [32]. Since in the RFMD one--nucleon loop diagrams are defined by
the point--like vertices and the Green functions of free virtual nucleons
with constant masses, there is only a naive way to satisfy requirement of
{\it locality} of effective interactions through the formal application of
the long--wavelength approximation to the computation of one--nucleon loop
diagrams. This approximation implies the expansion of one--nucleon loop
diagrams in powers of external momenta by keeping only the leading terms of
the expansion. Of course, the application of such anapproximation to the
computation of one--nucleon loop diagrams, when on--mass shell the energy
of the deuteron exceeds twice the masses of virtual nucleons, can seem
rather unjustified.

However, in this connection we would like to recall that the analogous
problem encounters itself for the derivation of Effective Chiral
Lagrangians [28,31] within effective quark models motivated by QCD like the
extended Nambu--Jona--Lasino (ENJL) model with chiral $U(3)\times U(3)$
symmetry [33--36]. Indeed, all phenomenological low--energy interactions
predicted by Effective Chiral Lagrangians [28,31] for the nonet of
low--lying vector mesons ($\rho(770)$, $\omega(780)$ and so  on) can be
derived within the ENJL model by calculating one--constituent quark loop
diagrams at leading order in the long--wavelength approximation. As has
turned out the long--wavelength approximation works very good in spite of
the fact that the constituent quark loop diagrams are defined by
point--like vertices of quark--meson interactions and the Green functions
of the free constituent quarks with constant masses $M_q \sim 330\,{\rm
MeV}$, and the masses of vector mesons exceed twice the constituent quark
mass. A formal justification of the validity of this approximation can be
given by attracting the Vector Dominance (VD) hypothesis [31,37] due to
which {\it the effective vertices of low--energy interactions of low--lying
vector mesons should be smooth functions of squared 4--momenta of
interacting mesons varying from on--mass shell to zero values}.  This
allows to calculate the vertices of low--energy interactions of vector
mesons keeping them off--mass shell around zero values of their squared
momenta [37], and then having had kept the leading terms of the
long--wavelength expansion to continue the resultant expression on--mass
shell. Such a procedure describes perfectly well all phenomenological
vertices of low--energy interactions of low--lying vector mesons predicted
by Effective Chiral Lagrangians [31,34--37].

One cannot say exactly, whether we really have in the RFMD some kind of the
VD hypothesis, i.e., smooth dependence of effective low--energy
interactions of the deuteron coupled to other particles on squared
4--momenta of interacting external particles including the deuteron.
However, the application of the long--wavelength approximation to the
computation of one--nucleon loop diagrams leads eventually to effective
local Lagrangians  describing reasonably well a dynamics of strong
low--energy nuclear interactions. The static parameters of the deuteron and
amplitudes of strong low--energy interactions of the deuteron coupled to
nucleons and other particles can be described in the RFMD in complete
agreement with the philosophy and technique of the derivation of Effective
Chiral Lagrangians within effective quark models motivated by QCD.

The agreement between the reaction rates for the neutron--proton radiative
capture, which is the M1 transition, calculated in the RFMD and the PMA is
not surprising. Indeed, it is known from particle physics that the
radiative decays of pseudoscalar and vector mesons like $\pi^0 \to
\gamma\gamma$, $\omega \to \pi^0 \gamma$ and so on, caused by the M1
transitions, can be computed both in the non--relativistic quark model [38]
, which is some kind of the PMA, and in the Effective Chiral Lagrangian
approach. In the non--relativistic quark model the matrix elements of these
decays are given in terms of magnetic moments of constituent quarks
proportional to $1/M_q$, whereas in the Effective Chiral Lagrangian
approach they are defined by the axial  anomaly and proportional to
$1/F_{\pi}$, the inverse power of the PCAC constant $F_{\pi}= 92.4\,{\rm
MeV}$ [16,27,39]. Equating the matrix elements of these decays calculated
in the non--relativistic quark model and in the Effective Chiral Lagrangian
approach  one can express a constituent quark mass in terms of the PCAC
constant $F_{\pi}$ [39]. The estimated value of the constituent quark mass
$M_q \simeq 400\,{\rm MeV}$ is comparable with the values $M_q = 330\div
380\;{\rm MeV}$ accepted in the literature [38]. This testifies that both
the non--relativistic quark model and the Effective Chiral Lagrangian
approach describe equally well the dynamics of strong low--energy
interactions of low--lying mesons even if for the decays caused by the M1
transitions. Referring to this example the agreement between the reaction
rates for the neutron--proton radiative capture calculated in the RFMD and
in the PMA, respectively, is understandable. The computation of the
astrophysical factor for the solar proton burning and the disintegration of
the deuteron by anti--neutrinos and photons in agreement with the PMA has
only confirmed a validity of dynamics of strong low--energy nuclear
interactions suggested by the RFMD.

The only problem which is left to discuss concerns the Coulomb repulsion
between protons in the process of the solar proton burning. As has been
shown by Kamionkowski and Bahcall [18] the Coulomb repulsion plays an
important role for calculation of the astrophysical factor $S_{\rm pp}(0) =
3.89 \times 10^{-25}\,{\rm MeV\,b}$. In our case the Coulomb repulsion is
taken into account in the form of the Gamow penetration factor $C(\eta)$
and, apart from weak interactions, only strong low--energy nuclear forces
are responsible for the value of the astrophysical factor $S_{\rm pp}(0) =
4.02 \times 10^{-25}\,{\rm MeV\,b}$. This discrepancy with the PMA, which
can be attributed to the peculiarity of the model with a local
four--nucleon interaction like Eq.~(\ref{label1.1}) and a description of
strong low--energy nuclear interactions through one--nucleon loop
exchanges, we are planning to resolve in our further development of the
RFMD.

Finally, we have also shown that in the RFMD with the effective local
four--nucleon interaction Eq.~(\ref{label1.1}) one can describe low--energy
elastic NN scattering in terms of the S--wave scattering length $a_{\rm
NN}$ and the effective range $r_{\rm NN}$ in spirit of the EFT approach
[11--13] and in complete agreement with low--energy nuclear phenomenology.

\section*{Appendix A. Computation of the amplitude of the solar proton burning}

In order to acquaint readers with the machinery of the RFMD we give below
the detailed derivation of the amplitude of the solar proton burning p + p
$\to$ D + e$^+$ + $\nu_{\rm e}$.

The process p + p $\to$ D + e$^+$ + $\nu_{\rm e}$ runs through the
intermediate W--boson exchange, i.e., p + p $\to$ D + W$^+$ $\to$ D + e$^+$
+ $\nu_{\rm e}$. The RFMD defines the transition in terms of the following
effective interactions [1,2] (see Eq.~(\ref{label1.1})):
$$
{\cal L}_{\rm npD}(x) = -ig_{\rm V}[\bar{p^c}(x)\gamma^{\mu}n(x) -
\bar{n^c}(x)\gamma^{\mu}p(x)]\,D^{\dagger}_{\mu}(x),
$$
$$
\hspace{-0.5in}{\cal L}^{\rm pp \to pp}_{\rm eff}(x) = \frac{1}{2}\,G_{\rm
\pi NN}\,\int d^3\rho \,\delta^{(3)}(\vec{\rho}\,)\,
$$
$$
\{[\bar{p}(t,\vec{x} + \frac{1}{2}\,\vec{\rho}\,)\,\gamma^{\mu}\gamma^5 p^c
(t,\vec{x} - \frac{1}{2}\,\vec{\rho}\,)]\,[\bar{p^c}(t,\vec{x} +
\frac{1}{2}\,\vec{\rho}\,)\,\gamma_{\mu}\gamma^5 p(t,\vec{x} -
\frac{1}{2}\,\vec{\rho}\,)]
$$
$$
 + [\bar{p}(t,\vec{x} + \frac{1}{2}\,\vec{\rho}\,)\,\gamma^5 p^c (t,\vec{x}
- \frac{1}{2}\,\vec{\rho}\,)]\,[\bar{p^c}(t,\vec{x} +
\frac{1}{2}\,\vec{\rho}\,)\,\gamma^5 p(t,\vec{x} -
\frac{1}{2}\,\vec{\rho}\,)]\},
$$
$$
\hspace{-0.5in}{\cal L}_{\rm npW}(x) = - \frac{g_{\rm
W}}{2\sqrt{2}}\,\cos\vartheta_C\,[\bar{n}(x)\gamma^{\nu}(1 - g_{\rm
A}\gamma^5) p(x)]\,W^-_{\nu}(x).\eqno({\rm A}.1)
$$
For convenience, in the effective local four--nucleon Lagrangian
Eq.~(\ref{label1.1}) we have introduced the interaction over a
radius--vector $\vec{\rho}$ of a relative movement of the protons with a
$\delta$--function $\delta^{(3)}(\vec{\rho}\,)$.

Then, the transition W$^+$ $\to$ e$^+$ + $\nu_{\rm e}$ is defined by the
Lagrangian
$$
{\cal L}_{\rm \nu_{\rm e} e^+ W}(x) = - \frac{g_{\rm
W}}{2\sqrt{2}}\,[\bar{\psi}_{\nu_{\rm e}}(x)\gamma^{\nu}(1 - \gamma^5)
\psi_{\rm e}(x)]\,W^+_{\nu}(x). \eqno({\rm A}.2)
$$
The electroweak coupling constant $g_{\rm W}$ is connected with the Fermi
weak constant $G_{\rm F}$ and the mass of the W--boson $M_{\rm W}$ through
the relation
$$
\frac{g^2_{\rm W}}{8M^2_{\rm W}} = \frac{G_{\rm F}}{\sqrt{2}}.\eqno({\rm A}.3)
$$
In order not to deal with the intermediate coupling constant $g_{\rm W}$ it
is convenient to apply to the computation of the matrix element of the
transition p + p $\to$ D + W$^+$ the interaction
$$
{\cal L}_{\rm npW}(x) = [\bar{n}(x)\gamma^{\nu}(1 - g_{\rm A}\gamma^5)
p(x)]\,W^-_{\nu}(x),\eqno({\rm A}.4)
$$
and for the description of the subsequent weak transition  W$^+$ $\to$
e$^+$ + $\nu_{\rm e}$ to replace the operator of the W--boson field by the
operator of the leptonic weak current
$$
W^-_{\nu}(x) \to -\frac{G_{\rm V}}{\sqrt{2}}\,[\bar{\psi}_{\nu_{\rm
e}}(x)\gamma_{\nu}(1 - \gamma^5) \psi_{\rm e}(x)].\eqno({\rm A}.5)
$$
The S matrix describing the transitions like  p + p $\to$ D + W$^+$ is defined
$$
{\rm S} = {\rm T}e^{\displaystyle i\int d^4x\,[{\cal L}_{\rm npD}(x)  +
{\cal L}_{\rm npW}(x) + {\cal L}^{\rm pp \to pp}_{\rm eff}(x) +
\ldots]},\eqno({\rm A}.6)
$$
where T is the time--ordering operator and the ellipses denote the
contribution of interactions irrelevant to the computation of the
transition p + p $\to$ D + W$^+$.

For the computation of the transition p + p $\to$ D + W$^+$ we have to
consider the third order term of the S matrix which reads
$$
{\rm S}^{(3)} = \frac{i^3}{3!}\int d^4x_1 d^4x_2 d^4x_3\,{\rm T}([{\cal
L}_{\rm npD}(x_1)  + {\cal L}_{\rm npW}(x_1) + {\cal L}^{\rm pp\to pp}_{\rm
eff}(x_1) + \ldots]
$$
$$
\times\,[{\cal L}_{\rm npD}(x_2)  + {\cal L}_{\rm npW}(x_2) + {\cal L}^{\rm
pp\to pp}_{\rm eff}(x_2) + \ldots]
$$
$$
\times\,[{\cal L}_{\rm npD}(x_3)  + {\cal L}_{\rm npW}(x_3) + {\cal L}^{\rm
pp\to pp}_{\rm eff}(x_3) + \ldots]) =
$$
$$
= -i \int d^4x_1 d^4x_2 d^4x_3\,{\rm T}({\cal L}^{\rm pp\to pp}_{\rm
eff}(x_1){\cal L}_{\rm npD}(x_2){\cal L}_{\rm npW}(x_3)) + \ldots
\eqno({\rm A}.7)
$$
The ellipses denote the terms which do not contribute to the matrix element
of the transition p + p $\to$ D + W$^+$ and the interaction ${\cal L}_{\rm
npW}(x)$ is given by Eq.~({\rm A}.4).  The S matrix element ${\rm
S}^{(3)}_{\rm pp\to DW^+}$ contributing to the transition p + p $\to$ D +
W$^+$ we determine as follows
$$
{\rm S}^{(3)}_{\rm pp\to DW^+} = -i \int d^4x_1 d^4x_2 d^4x_3\,{\rm
T}({\cal L}^{\rm pp\to pp}_{\rm eff}(x_1){\cal L}_{\rm npD}(x_2){\cal
L}_{\rm npW}(x_3)).\eqno({\rm A}.8)
$$
For the derivation of the effective Lagrangian ${\cal L}_{\rm pp\to
DW^+}(x)$ containing only the fields of the initial and the final particles
we should make all necessary contractions of the operators of the proton
and the neutron fields. These contractions we denote by the brackets as
$$
<{\rm S}^{(3)}_{\rm pp\to DW^+}> = -i \int d^4x_1 d^4x_2 d^4x_3\,<{\rm
T}({\cal L}^{\rm pp\to pp}_{\rm eff}(x_1){\cal L}_{\rm npD}(x_2){\cal
L}_{\rm npW}(x_3))>.\eqno({\rm A}.9)
$$
Now the effective Lagrangian ${\cal L}_{\rm pp\to DW^+}(x)$ related to the
S matrix element $<{\rm S}^{(3)}_{\rm pp\to DW^+}>$ can be defined as
$$
<{\rm S}^{(3)}_{\rm pp\to DW^+}> = i\int d^4x\,{\cal L}_{\rm pp\to DW^+}(x) =
$$
$$
= -i \int d^4x_1 d^4x_2 d^4x_3\,<{\rm T}({\cal L}^{\rm pp\to pp}_{\rm
eff}(x_1){\cal L}_{\rm npD}(x_2){\cal L}_{\rm npW}(x_3))>. \eqno({\rm A}.10)
$$
In terms of the operators of the interacting fields the effective
Lagrangian ${\cal L}_{\rm pp\to DW^+}(x)$ reads
$$
\int d^4x\,{\cal L}_{\rm pp\to DW^+}(x) =  - \int d^4x_1 d^4x_2
d^4x_3\,<{\rm T}({\cal L}^{\rm pp\to pp}_{\rm eff}(x_1){\cal L}_{\rm
npD}(x_2){\cal L}_{\rm npW}(x_3))>
$$
$$
= -\,\frac{1}{2}\,G_{\rm \pi NN}\,\times\,(-ig_{\rm V})\,\times \,(-g_{\rm
A})\int d^4x_1 d^4x_2 d^4x_3\,\int d^3\rho\,\delta^{(3)}(\vec{\rho}\,)\,
$$
$$
\times\,{\rm T}([\bar{p^c}(t_1,\vec{x}_1 +
\frac{1}{2}\,\vec{\rho}\,)\,\gamma_{\alpha}\gamma^5 p(t_1,\vec{x}_1 -
\frac{1}{2}\,\vec{\rho}\,)]\,D^{\dagger}_{\mu}(x_2)\,W^-_{\nu}(x_3))
$$
$$
\times <0|{\rm T}([\bar{p}(t_1,\vec{x}_1 +
\frac{1}{2}\,\vec{\rho}\,)\,\gamma^{\alpha}\gamma^5 p^c (t_1,\vec{x}_1 -
\frac{1}{2}\,\vec{\rho}\,)][\bar{p^c}(x_2)\gamma^{\mu}n(x_2) -
\bar{n^c}(x_2)\gamma^{\mu}p(x_2)]\,
$$
$$
\times\,[\bar{n}(x_3)\gamma^{\nu}\gamma^5 p(x_3)])|0>
-\,\frac{1}{2}\,G_{\rm \pi NN}\,\times\,(-ig_{\rm V})\,\times \,(-g_{\rm
A})\int d^4x_1 d^4x_2 d^4x_3\,\int d^3\rho \,\delta^{(3)}(\vec{\rho}\,)\,
$$
$$
\times\,{\rm T}([\bar{p^c}(t_1,\vec{x}_1 + \frac{1}{2}\,\vec{\rho}\,)
\gamma^5 p(t_1,\vec{x}_1 -
\frac{1}{2}\,\vec{\rho}\,)]\,D^{\dagger}_{\mu}(x_2)\,W^-_{\nu}(x_3))
$$
$$
\times <0|{\rm T}([\bar{p}(t_1,\vec{x}_1 + \frac{1}{2}\,\vec{\rho}\,)
\gamma^5 p^c (t_1,\vec{x}_1 -
\frac{1}{2}\,\vec{\rho}\,)][\bar{p^c}(x_2)\gamma^{\mu}n(x_2) -
\bar{n^c}(x_2)\gamma^{\mu}p(x_2)]\,
$$
$$
\times\,[\bar{n}(x_3)\gamma^{\nu}\gamma^5 p(x_3)])|0>.\eqno({\rm A}.11)
$$
Since p + p $\to$ D + W$^+$ is the Gamow--Teller transition, we have taken
into account the W--boson coupled with the axial  nucleon current.

Due to the relation $\bar{n^c}(x_2)\gamma^{\mu}p(x_2) = -
\bar{p^c}(x_2)\gamma^{\mu}n(x_2)$ the r.h.s. of Eq.~({\rm A}.11) can be
simplified as follows
$$
\int d^4x\,{\cal L}_{\rm pp\to DW^+}(x) =  - \int d^4x_1 d^4x_2
d^4x_3\,<{\rm T}({\cal L}^{\rm pp\to pp}_{\rm eff}(x_1){\cal L}_{\rm
npD}(x_2){\cal L}_{\rm npW}(x_3))>
$$
$$
=  G_{\rm \pi NN}\,\times\,(-ig_{\rm V})\,\times \,g_{\rm A}\int d^4x_1
d^4x_2 d^4x_3\,\int d^3\rho \,\delta^{(3)}(\vec{\rho}\,)\,
$$
$$
\times\,{\rm T}([\bar{p^c}(t_1,\vec{x}_1 +
\frac{1}{2}\,\vec{\rho}\,)\,\gamma_{\alpha}\gamma^5 p(t_1,\vec{x}_1 -
\frac{1}{2}\,\vec{\rho}\,)]\,D^{\dagger}_{\mu}(x_2)\,W^-_{\nu}(x_3))
$$
$$
\times<0|{\rm T}([\bar{p}(t_1,\vec{x}_1 +
\frac{1}{2}\,\vec{\rho}\,)\,\gamma^{\alpha}\gamma^5 p^c (t_1,\vec{x}_1 -
\frac{1}{2}\,\vec{\rho}\,)][\bar{p^c}(x_2)\gamma^{\mu}n(x_2)][\bar{n}(x_3)\gamma
^{\nu}\gamma^5 p(x_3)])|0>
$$
$$
 + G_{\rm \pi NN}\,\times\,(-ig_{\rm V})\,\times \,g_{\rm A}\int d^4x_1
d^4x_2 d^4x_3\,\int d^3\rho \,\delta^{(3)}(\vec{\rho}\,)\,
$$
$$
\times\,{\rm T}([\bar{p^c}(t_1,\vec{x}_1 + \frac{1}{2}\,\vec{\rho}\,)
\gamma^5 p(t_1,\vec{x}_1 -
\frac{1}{2}\,\vec{\rho}\,)]\,D^{\dagger}_{\mu}(x_2)\,W^-_{\nu}(x_3))
$$
$$
\times<0|{\rm T}([\bar{p}(t_1,\vec{x}_1 + \frac{1}{2}\,\vec{\rho}\,)
\gamma^5 p^c (t_1,\vec{x}_1 -
\frac{1}{2}\,\vec{\rho}\,)][\bar{p^c}(x_2)\gamma^{\mu}n(x_2)][\bar{n}(x_3)\gamma
^{\nu}\gamma^5 p(x_3)])|0>.\eqno({\rm A}.12)
$$
Making the necessary contractions we arrive at the expression
$$
\int d^4x\,{\cal L}_{\rm pp\to DW^+}(x) =  - \int d^4x_1 d^4x_2
d^4x_3\,<{\rm T}({\cal L}^{\rm pp\to pp}_{\rm eff}(x_1){\cal L}_{\rm
npD}(x_2){\cal L}_{\rm npW}(x_3))>
$$
$$
= 2\,\times\,G_{\rm \pi NN}\,\times\,(-ig_{\rm V})\,\times \,g_{\rm A}\int
d^4x_1 d^4x_2 d^4x_3\,\int d^3\rho \,\delta^{(3)}(\vec{\rho}\,)\,
$$
$$
\times\,{\rm T}([\bar{p^c}(t_1,\vec{x}_1 +
\frac{1}{2}\,\vec{\rho}\,)\,\gamma_{\alpha}\gamma^5 p(t_1,\vec{x}_1 -
\frac{1}{2}\,\vec{\rho}\,)]\,D^{\dagger}_{\mu}(x_2)\,W^-_{\nu}(x_3))
$$
$$
\times\,(-1)\,{\rm tr}\{\gamma^{\alpha}\gamma^5 (-i) S^c_F(t_1 -
t_2,\vec{x}_1 - \vec{x}_2 - \frac{1}{2}\,\vec{\rho}\,) \gamma^{\mu} (-i)
S_F(x_2 - x_3) \gamma^{\nu}\gamma^5
$$
$$
\times\,(-i) S_F(t_3 - t_1, \vec{x}_3  - \vec{x}_1 -
\frac{1}{2}\,\vec{\rho}\,)\}
$$
$$
+ 2\,\times\,G_{\rm \pi NN}\,\times\,(-ig_{\rm V})\,\times \,g_{\rm A}\int
d^4x_1 d^4x_2 d^4x_3\,\int d^3\rho \,\delta^{(3)}(\vec{\rho}\,)\,
$$
$$
\times\,{\rm T}([\bar{p^c}(t_1,\vec{x}_1 + \frac{1}{2}\,\vec{\rho}\,)
\gamma^5 p(t_1,\vec{x}_1 -
\frac{1}{2}\,\vec{\rho}\,)]\,D^{\dagger}_{\mu}(x_2)\,W^-_{\nu}(x_3))
$$
$$
\times\,(-1)\,{\rm tr}\{\gamma^5 (-i) S^c_F(t_1 - t_2,\vec{x}_1 - \vec{x}_2
- \frac{1}{2}\,\vec{\rho}\,) \gamma^{\mu} (-i) S_F(x_2 - x_3)
\gamma^{\nu}\gamma^5
$$
$$
\times\,(-i) S_F(t_3 - t_1, \vec{x}_3  - \vec{x}_1 -
\frac{1}{2}\,\vec{\rho}\,)\},\eqno({\rm A}.13)
$$
where the combinatorial factor 2 takes into account the fact that the
protons are identical particles in the nucleon loop.  Let us corroborate
the appearance of the factor 2 by a direct calculation:
$$
\int d^4x\,{\cal L}_{\rm pp\to DW^+}(x) =  - \int d^4x_1 d^4x_2
d^4x_3\,<{\rm T}({\cal L}^{\rm pp\to pp}_{\rm eff}(x_1){\cal L}_{\rm
npD}(x_2){\cal L}_{\rm npW}(x_3))>
$$
$$
= G_{\rm \pi NN}\,\times\,(-ig_{\rm V})\,\times \,g_{\rm A}\int d^4x_1
d^4x_2 d^4x_3\,\int d^3\rho \,\delta^{(3)}(\vec{\rho}\,)\,
$$
$$
\times\,{\rm T}([\bar{p^c}(t_1,\vec{x}_1 +
\frac{1}{2}\,\vec{\rho}\,)\,\gamma_{\alpha}\gamma^5 p(t_1,\vec{x}_1 -
\frac{1}{2}\,\vec{\rho}\,)]\,D^{\dagger}_{\mu}(x_2)\,W^-_{\nu}(x_3))
$$
$$
\times\,<0|{\rm T}([\bar{p}_{\alpha_1}(t_1,\vec{x}_1 +
\frac{1}{2}\,\vec{\rho}\,)\,(\gamma^{\alpha}\gamma^5 C)_{\alpha_1\beta_1}
\bar{p}_{\beta_1}(t_1,\vec{x}_1 -
\frac{1}{2}\,\vec{\rho}\,)][p_{\alpha_2}(x_2)(C\gamma^{\mu})_{\alpha_2\beta_2}n_
{\beta_2}(x_2)]
$$
$$
\times\,[\bar{n}_{\alpha_3}(x_3)(\gamma^{\nu}\gamma^5)_{\alpha_3\beta_3}
p_{\beta_3}(x_3)])|0>
+ (\gamma_{\alpha}\gamma^5 \otimes \gamma^{\alpha}\gamma^5 \to \gamma^5
\otimes \gamma^5) =
$$
$$
= G_{\rm \pi NN}\,\times\,(-ig_{\rm V})\,\times \,g_{\rm A}\int d^4x_1
d^4x_2 d^4x_3\,\int d^3\rho \,\delta^{(3)}(\vec{\rho}\,)\,
$$
$$
\times\,{\rm T}([\bar{p^c}(t_1,\vec{x}_1 +
\frac{1}{2}\,\vec{\rho}\,)\,\gamma_{\alpha}\gamma^5 p(t_1,\vec{x}_1 -
\frac{1}{2}\,\vec{\rho}\,)]\,D^{\dagger}_{\mu}(x_2)\,W^-_{\nu}(x_3))
$$
$$
\times\,\Big\{(\gamma^{\alpha}\gamma^5 C)_{\alpha_1\beta_1} (-i) S_F(t_2 -
t_1,\vec{x}_2 - \vec{x}_1 +
\frac{1}{2}\vec{\rho}\,)_{\alpha_2\beta_1}(C\gamma^{\mu})_{\alpha_2\beta_2}
(-i) S_F(x_2 - x_3)_{\beta_2\alpha_3}
$$
$$
\times\,(\gamma^{\nu}\gamma^5)_{\alpha_3\beta_3} (-i) S_F(t_3 -
t_1,\vec{x}_3 - \vec{x}_1 - \frac{1}{2}\vec{\rho}\,)_{\beta_3\alpha_1} -
$$
$$
- (-i) S_F(t_2 - t_1,\vec{x}_2 - \vec{x}_1 -
\frac{1}{2}\vec{\rho}\,)_{\alpha_2\alpha_1} (\gamma^{\alpha}\gamma^5
C)_{\alpha_1\beta_1} (C\gamma^{\mu})_{\alpha_2\beta_2} (-i) S_F(x_2 -
x_3)_{\beta_2\alpha_3}
$$
$$
\times\,(\gamma^{\nu}\gamma^5)_{\alpha_3\beta_3} (-i) S_F(t_3 -
t_1,\vec{x}_3 - \vec{x}_1 + \frac{1}{2}\vec{\rho}\,)_{\beta_3\beta_1}\Big\}
+ (\gamma_{\alpha}\gamma^5 \otimes \gamma^{\alpha}\gamma^5 \to \gamma^5
\otimes \gamma^5) =
$$
$$
= G_{\rm \pi NN}\,\times\,(-ig_{\rm V})\,\times \,g_{\rm A}\int d^4x_1
d^4x_2 d^4x_3\,\int d^3\rho \,\delta^{(3)}(\vec{\rho}\,)\,
$$
$$
\times\,{\rm T}([\bar{p^c}(t_1,\vec{x}_1 +
\frac{1}{2}\,\vec{\rho}\,)\,\gamma_{\alpha}\gamma^5 p(t_1,\vec{x}_1 -
\frac{1}{2}\,\vec{\rho}\,)]\,D^{\dagger}_{\mu}(x_2)\,W^-_{\nu}(x_3))
$$
$$
\times\,\Big\{{\rm tr}\{\gamma^{\alpha}\gamma^5 C (-i) S^T_F(t_2 -
t_1,\vec{x}_2 - \vec{x}_1 + \frac{1}{2}\vec{\rho}\,) C \gamma^{\mu} (-i)
S_F(x_2 - x_3) \gamma^{\nu}\gamma^5
$$
$$
\times\,(-i) S_F(t_3 - t_1,\vec{x}_3 - \vec{x}_1 - \frac{1}{2}\vec{\rho}\,)\}
$$
$$
- (-i) [S_F(t_2 - t_1,\vec{x}_2 - \vec{x}_1 - \frac{1}{2}\vec{\rho}\,)
\gamma^{\alpha}\gamma^5 C]_{\alpha_2\beta_1} [C \gamma^{\mu} (-i) S_F(x_2 -
x_3) \gamma^{\nu}\gamma^5
$$
$$
\times\,(-i) S_F(t_3 - t_1,\vec{x}_3 - \vec{x}_1 +
\frac{1}{2}\vec{\rho}\,)]_{\alpha_2\beta_1}\Big\}
+ (\gamma_{\alpha}\gamma^5 \otimes \gamma^{\alpha}\gamma^5 \to \gamma^5
\otimes \gamma^5) =
$$
$$
= G_{\rm \pi NN}\,\times\,(-ig_{\rm V})\,\times \,g_{\rm A}\int d^4x_1
d^4x_2 d^4x_3\,\int d^3\rho \,\delta^{(3)}(\vec{\rho}\,)\,
$$
$$
\times\,{\rm T}([\bar{p^c}(t_1,\vec{x}_1 +
\frac{1}{2}\,\vec{\rho}\,)\,\gamma_{\alpha}\gamma^5 p(t_1,\vec{x}_1 -
\frac{1}{2}\,\vec{\rho}\,)]\,D^{\dagger}_{\mu}(x_2)\,W^-_{\nu}(x_3))
$$
$$
\times\,\Big\{(-1) {\rm tr}\{\gamma^{\alpha}\gamma^5 [C^T S^T_F(t_2 -
t_1,\vec{x}_2 - \vec{x}_1 + \frac{1}{2}\vec{\rho}\,) C] \gamma^{\mu}(-i)
S_F(x_2 - x_3) \gamma^{\nu}\gamma^5
$$
$$
\times\,(-i) S_F(t_3 - t_1,\vec{x}_3 - \vec{x}_1 - \frac{1}{2}\vec{\rho}\,)\}
$$
$$
+ (-1) {\rm tr}\{(-i) [S^T_F(t_2 - t_1,\vec{x}_2 - \vec{x}_1 +
\frac{1}{2}\vec{\rho}\,) \gamma^{\alpha}\gamma^5 C]^T \gamma^{\mu}(-i)
S_F(x_2 - x_3) \gamma^{\nu}\gamma^5
$$
$$
\times\,(-i) S_F(t_3 - t_1,\vec{x}_3 - \vec{x}_1 +
\frac{1}{2}\vec{\rho}\,)\}\Big\}
+ (\gamma_{\alpha}\gamma^5 \otimes \gamma^{\alpha}\gamma^5 \to \gamma^5
\otimes \gamma^5) =
$$
$$
= G_{\rm \pi NN}\,\times\,(-ig_{\rm V})\,\times \,g_{\rm A}\int d^4x_1
d^4x_2 d^4x_3\,\int d^3\rho \,\delta^{(3)}(\vec{\rho}\,)\,
$$
$$
\times\,{\rm T}([\bar{p^c}(t_1,\vec{x}_1 +
\frac{1}{2}\,\vec{\rho}\,)\,\gamma_{\alpha}\gamma^5 p(t_1,\vec{x}_1 -
\frac{1}{2}\,\vec{\rho}\,)]\,D^{\dagger}_{\mu}(x_2)\,W^-_{\nu}(x_3))
$$
$$
\times\,\Big\{(-1) {\rm tr}\{\gamma^{\alpha}\gamma^5 (-i) S^c_F(t_1 -
t_2,\vec{x}_1 - \vec{x}_2 - \frac{1}{2}\vec{\rho}\,) \gamma^{\mu} (-i)
S_F(x_2 - x_3) \gamma^{\nu}\gamma^5
$$
$$
\times\,(-i) S_F(t_3 - t_1,\vec{x}_3 - \vec{x}_1 - \frac{1}{2}\vec{\rho}\,)\}
$$
$$
+ (-1) {\rm tr}\{ C^T (\gamma^{\alpha}\gamma^5)^T C (-i) S^c_F(t_1 -
t_2,\vec{x}_1 - \vec{x}_2 + \frac{1}{2}\vec{\rho}\,) \gamma^{\mu} (-i)
S_F(x_2 - x_3) \gamma^{\nu}\gamma^5
$$
$$
\times\,(-i) S_F(t_3 - t_1,\vec{x}_3 + \vec{x}_1 +
\frac{1}{2}\vec{\rho}\,)\}\Big\}
+ (\gamma_{\alpha}\gamma^5 \otimes \gamma^{\alpha}\gamma^5 \to \gamma^5
\otimes \gamma^5).\eqno({\rm A}.14)
$$
Here we have used the relation $C = -C^T$. Then, by applying the relation
$C^T (\gamma^{\alpha}\gamma^5)^T C = \gamma^{\alpha}\gamma^5$ we obtain the
following expression
$$
\int d^4x\,{\cal L}_{\rm pp\to DW^+}(x) =  - \int d^4x_1 d^4x_2
d^4x_3\,<{\rm T}({\cal L}^{\rm pp\to pp}_{\rm eff}(x_1){\cal L}_{\rm
npD}(x_2){\cal L}_{\rm npW}(x_3))>
$$
$$
= G_{\rm \pi NN}\,\times\,(-ig_{\rm V})\,\times \,g_{\rm A}\int d^4x_1
d^4x_2 d^4x_3\,\int d^3\rho \,\delta^{(3)}(\vec{\rho}\,)\,
$$
$$
\times\,\Big\{{\rm T}([\bar{p^c}(t_1,\vec{x}_1 +
\frac{1}{2}\,\vec{\rho}\,)\,\gamma_{\alpha}\gamma^5 p(t_1,\vec{x}_1 -
\frac{1}{2}\,\vec{\rho}\,)]\,D^{\dagger}_{\mu}(x_2)\,W^-_{\nu}(x_3))
$$
$$
\times\,(-1) {\rm tr}\{\gamma^{\alpha}\gamma^5 (-i) S^c_F(t_1 -
t_2,\vec{x}_1 - \vec{x}_2 - \frac{1}{2}\vec{\rho}\,) \gamma^{\mu} (-i)
S_F(x_2 - x_3) \gamma^{\nu}\gamma^5
$$
$$
\times\,(-i) S_F(t_3 - t_1,\vec{x}_3 - \vec{x}_1 - \frac{1}{2}\vec{\rho}\,)\}
$$
$$
+ {\rm T}([\bar{p^c}(t_1,\vec{x}_1 +
\frac{1}{2}\,\vec{\rho}\,)\,\gamma_{\alpha}\gamma^5 p(t_1,\vec{x}_1 -
\frac{1}{2}\,\vec{\rho}\,)]\,D^{\dagger}_{\mu}(x_2)\,W^-_{\nu}(x_3))
$$
$$
\times\,(-1) {\rm tr}\{ \gamma^{\alpha}\gamma^5 (-i) S^c_F(t_1 -
t_2,\vec{x}_1 - \vec{x}_2 + \frac{1}{2}\vec{\rho}\,) C \gamma^{\mu} (-i)
S_F(x_2 - x_3) \gamma^{\nu}\gamma^5
$$
$$
\times\,(-i) S_F(t_3 - t_1,\vec{x}_3 + \vec{x}_1 +
\frac{1}{2}\vec{\rho}\,)\}\Big\}
+ (\gamma_{\alpha}\gamma^5 \otimes \gamma^{\alpha}\gamma^5 \to \gamma^5
\otimes \gamma^5).\eqno({\rm A}.15)
$$
Using the property of the operators
$$
[\bar{p^c}(t_1,\vec{x}_1 + \frac{1}{2}\,\vec{\rho}\,)\,\Gamma
p(t_1,\vec{x}_1 - \frac{1}{2}\,\vec{\rho}\,)] = [\bar{p^c}(t_1,\vec{x}_1 -
\frac{1}{2}\,\vec{\rho}\,)\,\Gamma p(t_1,\vec{x}_1 +
\frac{1}{2}\,\vec{\rho}\,)] \eqno({\rm A}.16)
$$
for $\Gamma=\gamma^{\alpha}\gamma^5$ and $\gamma^5$, we get
$$
\int d^4x\,{\cal L}_{\rm pp\to DW^+}(x) =  - \int d^4x_1 d^4x_2
d^4x_3\,<{\rm T}({\cal L}^{\rm pp\to pp}_{\rm eff}(x_1){\cal L}_{\rm
npD}(x_2){\cal L}_{\rm npW}(x_3))>
$$
$$
= G_{\rm \pi NN}\,\times\,(-ig_{\rm V})\,\times \,g_{\rm A}\int d^4x_1
d^4x_2 d^4x_3\,\int d^3\rho \,\delta^{(3)}(\vec{\rho}\,)\,
$$
$$
\times\,\Big\{{\rm T}([\bar{p^c}(t_1,\vec{x}_1 +
\frac{1}{2}\,\vec{\rho}\,)\,\gamma_{\alpha}\gamma^5 p(t_1,\vec{x}_1 -
\frac{1}{2}\,\vec{\rho}\,)]\,D^{\dagger}_{\mu}(x_2)\,W^-_{\nu}(x_3))
$$
$$
\times\,(-1) {\rm tr}\{\gamma^{\alpha}\gamma^5 S^c_F(t_1 - t_2,\vec{x}_1 -
\vec{x}_2 - \frac{1}{2}\vec{\rho}\,) \gamma^{\mu} (-i) S_F(x_2 - x_3)
\gamma^{\nu}\gamma^5
$$
$$
\times\,(-i) S_F(t_3 - t_1,\vec{x}_3 - \vec{x}_1 - \frac{1}{2}\vec{\rho}\,)\}
$$
$$
+ {\rm T}([\bar{p^c}(t_1,\vec{x}_1 -
\frac{1}{2}\,\vec{\rho}\,)\,\gamma_{\alpha}\gamma^5 p(t_1,\vec{x}_1 +
\frac{1}{2}\,\vec{\rho}\,)]\,D^{\dagger}_{\mu}(x_2)\,W^-_{\nu}(x_3))
$$
$$
\times\,(-1) {\rm tr}\{(-i) \gamma^{\alpha}\gamma^5 S^c_F(t_1 -
t_2,\vec{x}_1 - \vec{x}_2 + \frac{1}{2}\vec{\rho}\,) C \gamma^{\mu} (-i)
S_F(x_2 - x_3) \gamma^{\nu}\gamma^5
$$
$$
\times\,(-i) S_F(t_3 - t_1,\vec{x}_3 + \vec{x}_1 +
\frac{1}{2}\vec{\rho}\,)\}\Big\}
+ (\gamma_{\alpha}\gamma^5 \otimes \gamma^{\alpha}\gamma^5 \to \gamma^5
\otimes \gamma^5).\eqno({\rm A}.17)
$$
Making a change of variables $\vec{\rho}\, \to - \vec{\rho}\,$ in the last
term, we arrive at the expression Eq.~({\rm A}.13).

Then, $S^c_F(x)$ and $S_F(x)$ are the Green functions of the free
anti--nucleon and nucleon field, respectively:
$$
S^c_F(x) = CS^T_F(-x) C^T = S_F(x) =
\int\frac{d^4k}{(2\pi)^4}\frac{\displaystyle e^{\displaystyle - i k\cdot
x}}{M_{\rm N} - \hat{k}}.\eqno({\rm A}.18)
$$
Passing to the momentum representation of the Green functions we get
$$
\int d^4x\,{\cal L}_{\rm pp\to DW^+}(x) =
$$
$$
= - i\,g_{\rm A}G_{\rm \pi NN}\frac{g_{\rm V}}{8\pi^2}\int d^4x_1 \int
\frac{d^4x_2 d^4k_2}{(2\pi)^4}\frac{d^4x_3
d^4k_3}{(2\pi)^4}\,e^{\displaystyle -i k_2\cdot (x_2-x_1)}e^{\displaystyle
-i k_3\cdot (x_3-x_1)}
$$
$$
\times\,\int d^3\rho \,\delta^{(3)}(\vec{\rho}\,)\,{\rm
T}([\bar{p^c}(t_1,\vec{x}_1 +
\frac{1}{2}\,\vec{\rho}\,)\,\gamma_{\alpha}\gamma^5 p(t_1,\vec{x}_1 -
\frac{1}{2}\,\vec{\rho}\,)]\,D^{\dagger}_{\mu}(x_2)\,W^-_{\nu}(x_3))
$$
$$
\times\,\int\frac{d^4k_1}{\pi^2i}\,e^{\displaystyle i\vec{q}\cdot
\vec{\rho}}\,{\rm tr}\Bigg\{\gamma^{\alpha}\gamma^5\frac{1}{M_{\rm N} -
\hat{k}_1 + \hat{k}_2}\gamma^{\mu}\frac{1}{M_{\rm N} -
\hat{k}_1}\gamma^{\nu}\gamma^5 \frac{1}{M_{\rm N} - \hat{k}_1 -
\hat{k}_3}\Bigg\}
$$
$$
- i\,g_{\rm A}G_{\rm \pi NN}\frac{g_{\rm V}}{8\pi^2}\int d^4x_1 \int
\frac{d^4x_2 d^4k_2}{(2\pi)^4}\frac{d^4x_3
d^4k_3}{(2\pi)^4}\,e^{\displaystyle -i k_2\cdot (x_2-x_1)}e^{\displaystyle
-i k_3\cdot (x_3-x_1)}
$$
$$
\times\,\int d^3\rho \,\delta^{(3)}(\vec{\rho}\,)\,{\rm
T}([\bar{p^c}(t_1,\vec{x}_1 + \frac{1}{2}\,\vec{\rho}\,)\,\gamma^5
p(t_1,\vec{x}_1 - \frac{1}{2}\,\vec{\rho}\,)]\,
D^{\dagger}_{\mu}(x_2)\,W^-_{\nu}(x_3))
$$
$$
\times\,\int\frac{d^4k_1}{\pi^2i}\,e^{\displaystyle i\vec{q}\cdot
\vec{\rho}}\,{\rm tr}\Bigg\{\gamma^5\frac{1}{M_{\rm N} - \hat{k}_1 +
\hat{k}_2}\gamma^{\mu}\frac{1}{M_{\rm N} - \hat{k}_1}\gamma^{\nu}\gamma^5
\frac{1}{M_{\rm N} - \hat{k}_1 - \hat{k}_3}\Bigg\},\eqno({\rm A}.19)
$$
where $\vec{q} = \vec{k}_1 + (\vec{k}_3 - \vec{k}_2)/2$.

In order to obtain the effective Lagrangian describing the process p + p
$\to$ D + e$^+$ + $\nu_{\rm e}$ we have to replace the operator of the
W--boson field by the operator of the leptonic weak current Eq.~({\rm A}.5):
$$
\int d^4x\,{\cal L}_{\rm pp\to D e^+ \nu_{\rm e}}(x) = $$
$$
= i\,g_{\rm A}G_{\rm \pi NN}\frac{G_{\rm V}}{\sqrt{2}} \frac{g_{\rm
V}}{8\pi^2}\int d^4x_1 \int \frac{d^4x_2 d^4k_2}{(2\pi)^4}\frac{d^4x_3
d^4k_3}{(2\pi)^4}\,e^{\displaystyle -i k_2\cdot (x_2-x_1)}e^{\displaystyle
-i k_3\cdot (x_3-x_1)}
$$
$$
\times\,\int d^3\rho \,\delta^{(3)}(\vec{\rho}\,)\,{\rm
T}([\bar{p^c}(t_1,\vec{x}_1 +
\frac{1}{2}\,\vec{\rho}\,)\,\gamma_{\alpha}\gamma^5 p(t_1,\vec{x}_1 -
\frac{1}{2}\,\vec{\rho}\,)]\,D^{\dagger}_{\mu}(x_2)\,[\bar{\psi}_{\nu_{\rm
e}}(x_3)\gamma_{\nu}(1 - \gamma^5) \psi_{\rm e}(x_3)])
$$
$$
\times\,\int\frac{d^4k_1}{\pi^2i}\,e^{\displaystyle i\vec{q}\cdot
\vec{\rho}}\,{\rm tr}\Bigg\{\gamma^{\alpha}\gamma^5\frac{1}{M_{\rm N} -
\hat{k}_1 + \hat{k}_2}\gamma^{\mu}\frac{1}{M_{\rm N} -
\hat{k}_1}\gamma^{\nu}\gamma^5 \frac{1}{M_{\rm N} - \hat{k}_1 -
\hat{k}_3}\Bigg\}
$$
$$
+ i\,g_{\rm A}G_{\rm \pi NN}\frac{G_{\rm V}}{\sqrt{2}}\frac{g_{\rm
V}}{8\pi^2}\int d^4x_1 \int \frac{d^4x_2 d^4k_2}{(2\pi)^4}\frac{d^4x_3
d^4k_3}{(2\pi)^4}\,e^{\displaystyle -i k_2\cdot (x_2-x_1)}e^{\displaystyle
-i k_3\cdot (x_3-x_1)}
$$
$$
\times\,\int d^3\rho \,\delta^{(3)}(\vec{\rho}\,)\,{\rm
T}([\bar{p^c}(t_1,\vec{x}_1 + \frac{1}{2}\,\vec{\rho}\,)\,\gamma^5
p(t_1,\vec{x}_1 - \frac{1}{2}\,\vec{\rho}\,)]\,
D^{\dagger}_{\mu}(x_2)\,[\bar{\psi}_{\nu_{\rm e}}(x_3)\gamma_{\nu}(1 -
\gamma^5) \psi_{\rm e}(x_3)])
$$
$$
\times\,\int\frac{d^4k_1}{\pi^2i}\,e^{\displaystyle i\vec{q}\cdot
\vec{\rho}}\,{\rm tr}\Bigg\{\gamma^5\frac{1}{M_{\rm N} - \hat{k}_1 +
\hat{k}_2}\gamma^{\mu}\frac{1}{M_{\rm N} - \hat{k}_1}\gamma^{\nu}\gamma^5
\frac{1}{M_{\rm N} - \hat{k}_1 - \hat{k}_3}\Bigg\}.\eqno({\rm A}.20)
$$
Now we are able to determine the matrix element of the  process p + p $\to$
D + e$^+$ + $\nu_{\rm e}$ as
$$
\int d^4x\,<{\rm D}(k_{\rm D}){\rm e}^+(k_{\rm e^+})\nu_{\rm e}(k_{\nu_{\rm
e}})|{\cal L}_{\rm pp\to D e^+ \nu_{\rm e}}(x)|p(p_1)p(p_2)> =
$$
$$
=(2\pi)^4\delta^{(4)}(k_{\rm D} + k_{\ell} - p_1 - p_2)\,\frac{{\cal
M}({\rm p} + {\rm p} \to {\rm D} + {\rm e}^+ + \nu_{\rm e})}{\displaystyle
\sqrt{2E_1V\,2E_2V\,2E_{\rm D}V\,2E_{\rm e^+}V\,2E_{\nu_{\rm e}}V}},
\eqno({\rm A}.21)
$$
where $k_{\ell} = k_{\rm e^+} + k_{\nu_{\rm e}}$ is the 4--momentum of the
leptonic pair, $E_i\,(i =1,2,{\rm D},{\rm e},\nu_{\rm e})$ are the energies
of the protons, the deuteron, positron and neutrino, $V$ is the
normalization volume.

Taking the r.h.s. of Eq.~({\rm A}.20) between the wave functions of the
initial $|p(p_1)p(p_2)>$ and the final $<{\rm D}(k_{\rm D}){\rm e}^+(k_{\rm
e^+})\nu_{\rm e}(k_{\nu_{\rm e}})|$ states we get
$$
(2\pi)^4\delta^{(4)}(k_{\rm D} + k_{\ell} - p_1 - p_2)\,\frac{{\cal M}({\rm
p} + {\rm p} \to {\rm D} + {\rm e}^+ + \nu_{e})}{\displaystyle
\sqrt{2E_1V\,2E_2V\,2E_{\rm D}V\,2E_{\rm e^+}V\,2E_{\nu_{\rm e}}V}}=
$$
$$
=  i\,g_{\rm A}G_{\rm \pi NN}\frac{G_{\rm V}}{\sqrt{2}} \frac{g_{\rm
V}}{8\pi^2}\int d^4x_1 \int \frac{d^4x_2 d^4k_2}{(2\pi)^4}\frac{d^4x_3
d^4k_3}{(2\pi)^4}\,e^{\displaystyle -i k_2\cdot (x_2-x_1)}e^{\displaystyle
-i k_3\cdot (x_3-x_1)}
$$
$$
\times\,\int d^3\rho \,\delta^{(3)}(\vec{\rho}\,)\,<{\rm D}(k_{\rm D}){\rm
e}^+(k_{\rm e^+})\nu_{\rm e}(k_{\nu_{\rm e}})|{\rm
T}([\bar{p^c}(t_1,\vec{x}_1 +
\frac{1}{2}\,\vec{\rho}\,)\,\gamma_{\alpha}\gamma^5 p(t_1,\vec{x}_1 -
\frac{1}{2}\,\vec{\rho}\,)]\,D^{\dagger}_{\mu}(x_2)\,
$$
$$
\times\,[\bar{\psi}_{\nu_{\rm e}}(x_3)\gamma_{\nu}(1 - \gamma^5) \psi_{\rm
e}(x_3)])|p(p_1) p(p_2)>
$$
$$
\times\,\int\frac{d^4k_1}{\pi^2i}\,e^{\displaystyle i\vec{q}\cdot
\vec{\rho}}\,{\rm tr}\Bigg\{\gamma^{\alpha}\gamma^5\frac{1}{M_{\rm N} -
\hat{k}_1 + \hat{k}_2}\gamma^{\mu}\frac{1}{M_{\rm N} -
\hat{k}_1}\gamma^{\nu}\gamma^5 \frac{1}{M_{\rm N} - \hat{k}_1 -
\hat{k}_3}\Bigg\}
$$
$$
+ i\,g_{\rm A}G_{\rm \pi NN}\frac{G_{\rm V}}{\sqrt{2}}\frac{g_{\rm
V}}{8\pi^2}\int d^4x_1 \int \frac{d^4x_2 d^4k_2}{(2\pi)^4}\frac{d^4x_3
d^4k_3}{(2\pi)^4}\,e^{\displaystyle -i k_2\cdot (x_2-x_1)}e^{\displaystyle
-i k_3\cdot (x_3-x_1)}
$$
$$
\times\,\int d^3\rho \,\delta^{(3)}(\vec{\rho}\,)\,<{\rm D}(k_{\rm D}){\rm
e}^+(k_{\rm e^+})\nu_{\rm e}(k_{\nu_{\rm e}})|{\rm
T}([\bar{p^c}(t_1,\vec{x}_1 + \frac{1}{2}\,\vec{\rho}\,)\,\gamma^5
p(t_1,\vec{x}_1 - \frac{1}{2}\,\vec{\rho}\,)]\, D^{\dagger}_{\mu}(x_2)\,
$$
$$
\times\,[\bar{\psi}_{\nu_{\rm e}}(x_3)\gamma_{\nu}(1 - \gamma^5) \psi_{\rm
e}(x_3)])|p(p_1) p(p_2)>
$$
$$
\times\,\int\frac{d^4k_1}{\pi^2i}\,e^{\displaystyle i\vec{q}\cdot
\vec{\rho}}\,{\rm tr}\Bigg\{\gamma^5\frac{1}{M_{\rm N} - \hat{k}_1 +
\hat{k}_2}\gamma^{\mu}\frac{1}{M_{\rm N} - \hat{k}_1}\gamma^{\nu}\gamma^5
\frac{1}{M_{\rm N} - \hat{k}_1 - \hat{k}_3}\Bigg\}.\eqno({\rm A}.22)
$$
Between the initial $|p(p_1)p(p_2)>$ and the final $<{\rm D}(k_{\rm D}){\rm
e}^+(k_{\rm e^+})\nu_{\rm e}(k_{\nu_{\rm e}})|$ states the matrix elements
are defined
$$
<{\rm D}(k_{\rm D}){\rm e}^+(k_{\rm e^+})\nu_{\rm e}(k_{\nu_{\rm e}})|{\rm
T}([\bar{p^c}(t_1,\vec{x}_1 +
\frac{1}{2}\,\vec{\rho}\,)\,\gamma_{\alpha}\gamma^5 p(t_1,\vec{x}_1 -
\frac{1}{2}\,\vec{\rho}\,)]\,D^{\dagger}_{\mu}(x_2)\,
$$
$$
\times\,[\bar{\psi}_{\nu_{\rm e}}(x_3)\gamma_{\nu}(1 - \gamma^5) \psi_{\rm
e}(x_3)])|p(p_1) p(p_2)>= [\bar{u^c}(p_2)\gamma_{\alpha}\gamma^5
u(p_1)][\bar{u}(k_{\nu_{\rm e}})\gamma_{\nu}
(1-\gamma^5)v(k_{\rm e^+})]\,
$$
$$
\times\,e^*_{\mu}(k_{\rm D})\,\sqrt{2}\,\psi_{\rm pp}(\vec{\rho})_{\rm
in}\,\frac{\displaystyle e^{\displaystyle -i(p_1+p_2)\cdot
x_1}\,e^{\displaystyle ik_{\rm D}\cdot x_2}\,e^{\displaystyle
ik_{{\ell}}\cdot x_3}}{\displaystyle \sqrt{2E_1V\,2E_2V\,2E_{\rm
D}V\,2E_{\rm e^+}V\,2E_{\nu_{\rm e}}V}},
$$
$$
<{\rm D}(k_{\rm D}){\rm e}^+(k_{\rm e^+})\nu_{\rm e}(k_{\nu_{\rm e}})|{\rm
T}([\bar{p^c}(t_1,\vec{x}_1 + \frac{1}{2}\,\vec{\rho}\,)\,\gamma^5
p(t_1,\vec{x}_1 - \frac{1}{2}\,\vec{\rho}\,)]\,D^{\dagger}_{\mu}(x_2)\,
$$
$$
\times\,[\bar{\psi}_{\nu_{\rm e}}(x_3)\gamma_{\nu}(1 - \gamma^5) \psi_{\rm
e}(x_3)])|p(p_1) p(p_2)>= [\bar{u^c}(p_2)\gamma^5
u(p_1)][\bar{u}(k_{\nu_{\rm e}})\gamma_{\nu}
(1-\gamma^5)v(k_{\rm e^+})]\,
$$
$$
\times\,e^*_{\mu}(k_{\rm D})\,\sqrt{2}\,\psi_{\rm pp}(\vec{\rho})_{\rm
in}\,\frac{\displaystyle e^{\displaystyle -i(p_1+p_2)\cdot
x_1}\,e^{\displaystyle ik_{\rm D}\cdot x_2}\,e^{\displaystyle
ik_{{\ell}}\cdot x_3}}{\displaystyle \sqrt{2E_1V\,2E_2V\,2E_{\rm
D}V\,2E_{\rm e^+}V\,2E_{\nu_{\rm e}}V}},\eqno({\rm A}.23)
$$
where $\psi_{\rm pp}(\vec{\rho})_{\rm in}$ is the wave function of the
relative movement of the free protons in the ${^1}{\rm S}_0$--state
normalized to unit density [40]:
$$
\psi_{\rm pp}(\vec{\rho})_{\rm in}= \frac{\sin k\rho}{k\rho},\eqno({\rm A}.24)
$$
where $k$ is a 3--momentum of a relative movement of the protons. Since the
spatial part of the wave function of the protons is symmetric under
permutations of the protons, so the spinorial part should be antisymmetric.
In our approach the spinorial part of the wave function of the protons is
described by $[\bar{u^c}(p_2)\gamma_{\alpha}\gamma^5 u(p_1)]$ and
$[\bar{u^c}(p_2)\gamma^5 u(p_1)]$, antisymmetric under permutations of the
protons: $[\bar{u^c}(p_2) \gamma_{\alpha} \gamma^5 u(p_1)] = -
[\bar{u^c}(p_1) \gamma_{\alpha} \gamma^5 u(p_2)]$ and $[\bar{u^c}(p_2)
\gamma^5  u(p_1)] = - [\bar{u^c}(p_1) \gamma^5 u(p_2)]$.

Now let us discuss in details the computation of the matrix elements:
$$
<0|\bar{p^c}(t_1,\vec{x}_1 + \frac{1}{2}\,\vec{\rho}\,)\,\Gamma\,
p(t_1,\vec{x}_1 - \frac{1}{2}\,\vec{\rho}\,)|p(p_1) p(p_2)>, \eqno({\rm
A}.25)
$$
where we have denoted $\Gamma = \gamma_{\alpha}\gamma^5$ or $\gamma^5$.

In the quantum field theory approach the wave function $|p(p_1) p(p_2)>$
should be described in terms of the operators of the creation of the
protons $a^{\dagger}(\vec{p}_1,\sigma_1)$ and
$a^{\dagger}(\vec{p}_2,\sigma_2)$, where $\vec{p}_i$ and
$\sigma_i\,(i=1,2)$ are the 3--momenta and the polarizations of the
protons. Therefore, $|p(p_1) p(p_2)>$ reads
$$
|p(p_1) p(p_2)> =\frac{1}{\sqrt{2}}
a^{\dagger}(\vec{p}_1,\sigma_1)\,a^{\dagger}(\vec{p}_2,\sigma_2)|0>.\eqno({\rm A
}.26)
$$
The wave function Eq.~({\rm A}.26) is taken in the standard form [41]. It
is antisymmetric under permutations of the protons due to the
anti--commutation relation
\[a^{\dagger}(\vec{p}_1,\sigma_1) \,a^{\dagger}(\vec{p}_2,\sigma_2) = -
a^{\dagger}(\vec{p}_2,\sigma_2) \,a^{\dagger}(\vec{p}_1,\sigma_1)\]
and normalized to unity. The factor $1/\sqrt{2}$ takes into account that
the protons are correlated in the initial state.

The operators of the proton fields $\bar{p^c}(t_1,\vec{x}_1 +
\frac{1}{2}\,\vec{\rho}\,)$ and $p(t_1,\vec{x}_1 -
\frac{1}{2}\,\vec{\rho}\,)$ we represent in terms of the plane--wave
expansions
$$
\bar{p^c}(t_1,\vec{x}_1 + \frac{1}{2}\,\vec{\rho}\,) =
\sum_{\vec{q}_1,\alpha_1}\frac{1}{\displaystyle
\sqrt{2E_{\vec{q}_1}V}}\Bigg[
a(\vec{q}_1,\alpha_1)\,\bar{u^c}(q_1)\,e^{\displaystyle -iE_{\vec{q}_1}t_1
+ i\vec{q}_1\cdot (\vec{x}_1 + \vec{\rho}/2)}
$$
$$
\hspace{0.5in}+
b^{\dagger}(\vec{q}_1,\alpha_1)\,\bar{v^c}(q_1)\,e^{\displaystyle
iE_{\vec{q}_1}t_1 - i\vec{q}_1\cdot (\vec{x}_1 + \vec{\rho}/2)}\Bigg],
$$
$$
p(t_1,\vec{x}_1 - \frac{1}{2}\,\vec{\rho}\,) =
\sum_{\vec{q}_2,\alpha_2}\frac{1}{\displaystyle
\sqrt{2E_{\vec{q}_2}V}}\Bigg[
a(\vec{q}_2,\alpha_2)\,u(q_2)\,e^{\displaystyle -iE_{\vec{q}_2}t_1 +
i\vec{q}_2\cdot (\vec{x}_1 - \vec{\rho}/2)}
$$
$$
 +  b^{\dagger}(\vec{q}_2,\alpha_2)\,v(q_2)\,e^{\displaystyle
iE_{\vec{q}_2}t_1 - i\vec{q}_2\cdot (\vec{x}_1 -
\vec{\rho}/2)}\Bigg],\eqno({\rm A}.27)
$$
where $a(\vec{q}_i,\alpha_i)\,(i=1,2)$ and
$b^{\dagger}(\vec{q}_i,\alpha_i)\,(i=1,2)$ are the operators of the
annihilation and the creation of protons and ani-protons, respectively. The
computation of the matrix element Eq.~({\rm A}.25) runs the following way.
Holding only the terms containing the operators of the annihilation of the
protons we get
$$
<0|\bar{p^c}(t_1,\vec{x}_1 + \frac{1}{2}\,\vec{\rho}\,)\,\Gamma
p(t_1,\vec{x}_1 - \frac{1}{2}\,\vec{\rho}\,)|p(p_1) p(p_2)> =
$$
$$
=\sum_{\vec{q}_1,\alpha_1}\sum_{\vec{q}_2,\alpha_2}\frac{1}{\displaystyle
\sqrt{2E_{\vec{q}_1}V}}\frac{1}{\displaystyle
\sqrt{2E_{\vec{q}_2}V}}\,e^{\displaystyle -i(q_1 + q_2)\cdot x_1 +
i(\vec{q}_1 - \vec{q}_2)\cdot \vec{\rho}/2}
$$
$$
\times\,[\bar{u^c}(q_1)
\,\Gamma\,u(q_2)]
\,\frac{1}{\sqrt{2}}
<0|a(\vec{q}_1,\alpha_1) \,
a(\vec{q}_2,\alpha_2) \,
a^{\dagger}(\vec{p}_1,\sigma_1)
\,a^{\dagger}(\vec{p}_2,\sigma_2)|0>.\eqno({\rm A}.28)
$$
The vacuum expectation value $<0|a(\vec{q}_1,\alpha_1)
\,a(\vec{q}_1,\alpha_1) \,a^{\dagger}(\vec{p}_1,\sigma_1)
\,a^{\dagger}(\vec{p}_2,\sigma_2)|0>$ reads:
$$
<0|a(\vec{q}_1,\alpha_1) \,
a(\vec{q}_1,\alpha_1) \,
a^{\dagger}(\vec{p}_1,\sigma_1) \,
a^{\dagger}(\vec{p}_2,\sigma_2)|0> =
$$
$$
=-\delta_{\vec{q}_1\vec{p}_1}\,
\delta_{\alpha_1\sigma_1}\,
\delta_{\vec{q}_2\vec{p}_2}\,
\delta_{\alpha_2\sigma_2} +
\delta_{\vec{q}_2\vec{p}_1}\,
\delta_{\alpha_2\sigma_1}\,
\delta_{\vec{q}_1\vec{p}_2}\,
\delta_{\alpha_1\sigma_2},\eqno({\rm A}.29)
$$
where we have used the anti--commutation relations
$$
a(\vec{q},\alpha) \,
a^{\dagger}(\vec{p},\sigma) + a^{\dagger}(\vec{p},
\sigma)a(\vec{q},\alpha) = \delta_{\vec{q}\vec{p}}
\,\delta_{\alpha\sigma}\eqno({\rm A}.30)
$$
and the properties of the operators of the creation and the annihilation:
$<0|a^{\dagger}(\vec{p},\sigma) = 0$ and $a(\vec{q},\alpha)|0> = 0$.

Substituting Eq.~({\rm A}.29) in Eq.~({\rm A}.28) and summing up the
momenta and the spinorial indices we arrive at the expression
$$
<0|\bar{p^c}(t_1,\vec{x}_1 + \frac{1}{2}\,\vec{\rho}\,)\,\Gamma
p(t_1,\vec{x}_1 - \frac{1}{2}\,\vec{\rho}\,)
|p(p_1) p(p_2)>
=-\frac{\displaystyle e^{\displaystyle -i(p_1 + p_2)\cdot
x_1}}{\displaystyle \sqrt{2E_1V\,2E_2V}}\,
$$
$$
\times\,\frac{1}{\sqrt{2}}
\,\Bigg([\bar{u^c}(p_1)
\,\Gamma\,u(p_2)]
\,e^{\displaystyle i(\vec{p}_1 - \vec{p}_2)\cdot \vec{\rho}/2} -
[\bar{u^c}(p_2)
\,\Gamma\,u(p_1)]
\,e^{\displaystyle - i(\vec{p}_1 - \vec{p}_2)\cdot \vec{\rho}/2}\Bigg) =
$$
$$
= \frac{\displaystyle e^{\displaystyle -i(p_1 + p_2)\cdot
x_1}}{\displaystyle
\sqrt{2E_1V\,2E_2V}}\,\sqrt{2}\,[\bar{u^c}(p_2)\,\Gamma\,u(p_1)]\,\frac{1}{2}\,\
Bigg(e^{\displaystyle i(\vec{p}_1 - \vec{p}_2)\cdot \vec{\rho}/2} +
e^{\displaystyle - i(\vec{p}_1 - \vec{p}_2)\cdot
\vec{\rho}/2}\Bigg),\eqno({\rm A}.31)
$$
where the relation $[\bar{u^c}(p_1)\,\Gamma\,u(p_2)] = -
[\bar{u^c}(p_2)\,\Gamma\,u(p_1)]$ has been used. The sum of the exponentials
$$
\frac{1}{2}\,\Bigg(e^{\displaystyle i(\vec{p}_1 - \vec{p}_2)\cdot
\vec{\rho}/2} +  e^{\displaystyle - i(\vec{p}_1 - \vec{p}_2)\cdot
\vec{\rho}/2}\Bigg) \eqno({\rm A}.32)
$$
describes the spatial part of the wave function of the relative movement of
the free protons. This wave function is  symmetric  under permutations of
the protons and normalized to unit density [40]. Since the protons should
be in the ${^1}{\rm S}_0$--state, expanding exponentials into spherical
harmonics and keeping only the S--wave contribution we obtain [40]:
$$
\frac{1}{2}\,\Bigg(e^{\displaystyle i \vec{k}\cdot \vec{\rho}} +
e^{\displaystyle -i \vec{k}\cdot \vec{\rho}}\Bigg) = \frac{\sin
k\rho}{k\rho}+ \ldots,\eqno({\rm A}.33)
$$
where  $\vec{k}=(\vec{p}_1 - \vec{p}_2)/2$ is the relative momentum of the
protons.  This completes the explanation of the derivation of the matrix
elements in Eq.~({\rm A}.23).

Substituting the matrix elements  Eq.~({\rm A}.23) in the r.h.s. of
Eq.~({\rm A}.22) we obtain the matrix element of the solar proton burning
in the following form
$$
(2\pi)^4\delta^{(4)}(k_{\rm D} + k_{\ell} - p_1 - p_2)\,i{\cal M}({\rm p} +
{\rm p} \to {\rm D} + {\rm e}^+ + \nu_{e}) = $$
$$
= - \sqrt{2}\,C(\eta)\,g_{\rm A}G_{\rm \pi NN}\frac{G_{\rm V}}{\sqrt{2}}
\frac{g_{\rm V}}{8\pi^2}[\bar{u^c}(p_2)\gamma_{\alpha}\gamma^5
u(p_1)][\bar{u}(k_{\nu_{\rm e}})\gamma_{\nu}(1-\gamma^5)v(k_{\rm
e^+})]\,e^*_{\mu}(k_{\rm D})
$$
$$
\times \int d^4x_1 \int \frac{d^4x_2 d^4k_2}{(2\pi)^4}\frac{d^4x_3
d^4k_3}{(2\pi)^4}\,e^{\displaystyle i (k_2 + k_3 - p_1 - p_2)\cdot
x_1}\,e^{\displaystyle i(k_{\rm D} - k_2)\cdot x_2}\,e^{\displaystyle
i(k_{\ell} - k_3)\cdot x_3}
$$
$$
\times\int d^3\rho   \delta^{(3)}(\vec{\rho}\,)\frac{\sin
k\rho}{k\rho}\int\frac{d^4k_1}{\pi^2i} e^{\displaystyle i\vec{q}\cdot
\vec{\rho}} {\rm tr}\Bigg\{\gamma^{\alpha}\gamma^5\frac{1}{M_{\rm N} -
\hat{k}_1 + \hat{k}_2}\gamma^{\mu}\frac{1}{M_{\rm N} -
\hat{k}_1}\gamma^{\nu}\gamma^5 \frac{1}{M_{\rm N} - \hat{k}_1 -
\hat{k}_3}\Bigg\}
$$
$$
-  \sqrt{2}\,C(\eta)\,g_{\rm A}G_{\rm \pi NN}\frac{G_{\rm V}}{\sqrt{2}}
\frac{g_{\rm V}}{8\pi^2}[\bar{u^c}(p_2)\gamma^5 u(p_1)][\bar{u}(k_{\nu_{\rm
e}})\gamma_{\nu}(1-\gamma^5)v(k_{\rm e^+})]\,e^*_{\mu}(k_{\rm D})
$$
$$
\times\int d^4x_1 \int \frac{d^4x_2 d^4k_2}{(2\pi)^4}\frac{d^4x_3
d^4k_3}{(2\pi)^4}\,e^{\displaystyle i (k_2 + k_3 - p_1 - p_2)\cdot
x_1}\,e^{\displaystyle i(k_{\rm D} - k_2)\cdot x_2}\,e^{\displaystyle
i(k_{\ell} - k_3)\cdot x_3}
$$
$$
\times\int d^3\rho \delta^{(3)}(\vec{\rho}\,)\frac{\sin k\rho}{k\rho}
\int\frac{d^4k_1}{\pi^2i}e^{\displaystyle i\vec{q}\cdot \vec{\rho}}{\rm
tr}\Bigg\{\gamma^5\frac{1}{M_{\rm N} - \hat{k}_1 +
\hat{k}_2}\gamma^{\mu}\frac{1}{M_{\rm N} - \hat{k}_1}\gamma^{\nu}\gamma^5
\frac{1}{M_{\rm N} - \hat{k}_1 - \hat{k}_3}\Bigg\},\eqno({\rm A}.35)
$$
where we have appended the Gamow penetration factor $C(\eta)$ taking into
account the Coulomb repulsion between the protons [2].

Integrating over $x_1$, $x_2$, $x_3$, $k_2$ and $k_3$ we obtain in the
r.h.s. of Eq.~({\rm A}.35) the $\delta$--function describing the
4--momentum conservation. Then, the matrix element of the  p + p $\to$ D +
e$^+$ + $\nu_{\rm e}$ process becomes equal
$$
i{\cal M}({\rm p} + {\rm p} \to {\rm D} + {\rm e}^+ + \nu_{e}) =
-\sqrt{2}\,C(\eta)\,g_{\rm A}G_{\rm \pi NN}\frac{G_{\rm V}}{\sqrt{2}}
\frac{g_{\rm V}}{8\pi^2}
$$
$$
\times\,[\bar{u^c}(p_2)\gamma_{\alpha}\gamma^5 u(p_1)][\bar{u}(k_{\nu_{\rm
e}})\gamma_{\nu}(1-\gamma^5)v(k_{\rm e^+})]\,e^*_{\mu}(k_{\rm D})
$$
$$
\times\,\int\frac{d^4k_1}{\pi^2i}\,{\rm
tr}\Bigg\{\gamma^{\alpha}\gamma^5\frac{1}{M_{\rm N} - \hat{k}_1 +
\hat{k}_2}\gamma^{\mu}\frac{1}{M_{\rm N} - \hat{k}_1}\gamma^{\nu}\gamma^5
\frac{1}{M_{\rm N} - \hat{k}_1 - \hat{k}_3}\Bigg\}
$$
$$
- \sqrt{2}\,C(\eta)\,g_{\rm A}G_{\rm \pi NN}\frac{G_{\rm V}}{\sqrt{2}}
\frac{g_{\rm V}}{8\pi^2}[\bar{u^c}(p_2)\gamma^5 u(p_1)][\bar{u}(k_{\nu_{\rm
e}})\gamma_{\nu}(1-\gamma^5)v(k_{\rm e^+})]\,e^*_{\mu}(k_{\rm D})
$$
$$
\times \int \frac{d^4k_1}{\pi^2i}\,{\rm tr} \Bigg\{ \gamma^5\frac{1}{M_{\rm
N} - \hat{k}_1 + \hat{k}_{\rm D}}\gamma^{\mu}\frac{1}{M_{\rm N} -
\hat{k}_1}\gamma^{\nu}\gamma^5 \frac{1}{M_{\rm N} - \hat{k}_1 -
\hat{k}_{\ell}}\Bigg\},\eqno({\rm A}.36)
$$
where we have integrated over a relative radius--vector $\vec{\rho}\,$ too.
It is convenient to represent the matrix element Eq.~({\rm A}.35) in terms
of the structure functions ${\cal J}^{\alpha\mu\nu}(k_{\rm D}, k_{\ell};
Q)$ and ${\cal J}^{\mu\nu}(k_{\rm D}, k_{\ell}; Q)$:
$$
i{\cal M}({\rm p} + {\rm p} \to {\rm D} + {\rm e}^+ + \nu_{e}) =
$$
$$
= - \,C(\eta)\,G_{\rm V}\,g_{\rm A}G_{\rm \pi NN}\frac{g_{\rm
V}}{8\pi^2}[\bar{u^c}(p_2)\gamma_{\alpha}\gamma^5
u(p_1)][\bar{u}(k_{\nu_{\rm e}})\gamma_{\nu}(1-\gamma^5)v(k_{\rm
e^+})]\,e^*_{\mu}(k_{\rm D})\,{\cal J}^{\alpha\mu\nu}(k_{\rm D}, k_{\ell};
Q)
$$
$$
- \,C(\eta)\,G_{\rm V}\,g_{\rm A}G_{\rm \pi NN} \frac{g_{\rm
V}}{8\pi^2}[\bar{u^c}(p_2)\gamma^5 u(p_1)][\bar{u}(k_{\nu_{\rm
e}})\gamma_{\nu}(1-\gamma^5)v(k_{\rm e^+})]\,e^*_{\mu}(k_{\rm D}){\cal
J}^{\mu\nu}(k_{\rm D}, k_{\ell}; Q),\eqno({\rm A}.37)
$$
where the structure functions ${\cal J}^{\alpha\mu\nu}(k_{\rm D}, k_{\ell};
Q)$ and ${\cal J}^{\mu\nu}(k_{\rm D}, k_{\ell}; Q)$ are defined as [2]
$$
{\cal J}^{\alpha\mu\nu}(k_{\rm D}, k_{\ell}; Q) =
$$
$$
=\int\frac{d^4k}{\pi^2i} \,{\rm tr}
\Bigg\{\gamma^{\alpha}\gamma^5\frac{1}{M_{\rm N} - \hat{k} - \hat{Q} +
\hat{k}_{\rm D}}\gamma^{\mu}\frac{1}{M_{\rm N} - \hat{k} -
\hat{Q}}\gamma^{\nu}\gamma^5 \frac{1}{M_{\rm N} - \hat{k} - \hat{Q} -
\hat{k}_{\ell}}\Bigg\},
$$
$$
{\cal J}^{\mu\nu}(k_{\rm D}, k_{\ell}; Q) =
$$
$$
=\int\frac{d^4k}{\pi^2i} \,{\rm tr}\Bigg\{\gamma^5\frac{1}{M_{\rm N} -
\hat{k} - \hat{Q} + \hat{k}_{\rm D}}\gamma^{\mu}\frac{1}{M_{\rm N} -
\hat{k} - \hat{Q}}\gamma^{\nu}\gamma^5 \frac{1}{M_{\rm N} - \hat{k} -
\hat{Q} - \hat{k}_{\ell}}\Bigg\}.\eqno({\rm A}.38)
$$
We have introduced a 4--vector $Q = a\,k_{\rm D} + b\,k_{\ell}$ caused by
an arbitrary shift of a virtual momentum  with arbitrary parameters $a$ and
$b$.

Thus, the problem of the computation of the matrix element of the p + p
$\to$ D + e$^+$ + $\nu_{\rm e}$ process reduces to the problem of the
computation of the structure functions Eq.~({\rm A}.38). Since the energy
of the leptonic pair is small compared with the nucleon mass, we can set in
the integrand $k^{\mu}_{\ell} = 0$ [2]. This gives
$$
{\cal J}^{\alpha\mu\nu}(k_{\rm D}, k_{\ell}; Q) =
$$
$$
=\int\frac{d^4k}{\pi^2i} \,{\rm tr}
\Bigg\{\gamma^{\alpha}\gamma^5\frac{1}{M_{\rm N} - \hat{k} - \hat{Q} +
\hat{k}_{\rm D}}\gamma^{\mu}\frac{1}{M_{\rm N} - \hat{k} -
\hat{Q}}\gamma^{\nu}\gamma^5 \frac{1}{M_{\rm N} - \hat{k} - \hat{Q}}\Bigg\},
$$
$$
{\cal J}^{\mu\nu}(k_{\rm D}, k_{\ell}; Q) =
$$
$$
=\int\frac{d^4k}{\pi^2i} \,{\rm tr}\Bigg\{\gamma^5\frac{1}{M_{\rm N} -
\hat{k} - \hat{Q} + \hat{k}_{\rm D}}\gamma^{\mu}\frac{1}{M_{\rm N} -
\hat{k} - \hat{Q}}\gamma^{\nu}\gamma^5 \frac{1}{M_{\rm N} - \hat{k} -
\hat{Q}}\Bigg\},\eqno({\rm A}.39)
$$
For the calculation of the momentum integrals we would follow the
philosophy of the derivation of Effective Chiral Lagrangians within
effective quark models motivated by QCD [33--36], in particularly, Chiral
perturbation theory at the quark level (CHPT)$_q$ [35] formulated on the
basis of the ENJL model induced by the effective low--energy QCD with
linearly rising confinement potential [42]. In (CHPT)$_q$ all low--energy
vertices of meson interactions are determined by constituent quark loop
diagrams with point--like quark--meson vertices and the Green functions of
the free constituent quarks with constant masses $M_q = 330\,{\rm MeV}$
[35]. To the computation of the momentum integrals one applies a
generalized hypothesis of Vector Dominance [31,37] postulating a smooth
dependence of low--energy vertices of hadron interactions on squared
4--momenta of interacting mesons. Due to this hypothesis one can hold all
external particles off--mass shell at squared 4--momenta $p^2$ much less
than $M^2_q$, i.e., $M^2_q\gg p^2$. Then, after the computation of the
momentum integrals at leading order in long--wavelength expansion, i.e., in
powers of external momenta, the resultant expression should be continued
on--mass shell of interacting particles. Within the framework of this
procedure one can restore completely all variety of phenomenological
vertices of low--energy meson interactions predicted by Effective Chiral
Lagrangians [28,31,34--36]. It is important to emphasize that this
procedure works good not only for light mesons like $\pi$--meson, which
mass is less than the mass of constituent quarks, but for vector mesons
like $\rho(770)$, $\omega(780)$ and so on, which masses are twice larger
than the constituent quark mass. Since the former resembles the RFMD, where
the mass of the deuteron amounts to twice the mass of virtual nucleons, we
expect that the long--wavelength approximation should work in the RFMD as
well as in the effective quark models with chiral $U(3) \times U(3)$
symmetry applied to the derivation of Effective Chiral Lagrangians.

Thus, for the computation of the momentum integrals we assume that the
deuteron is off--mass shell and $M_{\rm N} \gg \sqrt{k^2_{\rm D}}$. Then,
we expand the integrand of the structure functions Eq.~({\rm A}.39) in
powers of $k_{\rm D}$ keeping only the leading contributions. The result of
the computation we continue on--mass shell of the deuteron $k^2_{\rm D} \to
M^2_{\rm D}$ [2].

Keeping the leading terms of the expansion in powers of $k_{\rm D}$  we get
[2]:
$$
{\cal J}^{\alpha\mu\nu}(k_{\rm D}, k_{\ell}; Q) = 3\,(k^{\alpha}_{\rm D}
g^{\nu\mu} - k^{\nu}_{\rm D} g^{\mu\alpha}) +
\frac{1}{9}\,(1+2a)\,(k^{\alpha}_{\rm D} g^{\nu\mu}
+ k^{\nu}_{\rm D} g^{\mu\alpha}),
$$
$$
{\cal J}^{\mu\nu}(k_{\rm D}, k_{\ell}; Q) =  g^{\mu\nu} 4\,M_{\rm
N}\,J_2(M_{\rm N}), \eqno({\rm A}.40)
$$
where due to the relation $k_{\rm D}\cdot e^*_{\rm D}(k_{\rm D}) = 0$ the
terms proportional to $k^{\mu}_{\rm D}$ have been dropped out. Then,
$J_2(M_{\rm N})$ is a logarithmically divergent integral defined in the
RFMD in terms of the cut--off $\Lambda_D = 68.452\,{\rm MeV}$ such as
$\Lambda_{\rm D} \ll M_{\rm N}$ [1,2]:
$$
J_2(M_{\rm N}) =
 \int\frac{d^4k}{\pi^2i}
  \frac{1}{(M^2_{\rm N} - k^2)^2} = 2\int\limits^{\Lambda_{\rm D}}_0
\frac{d|\vec{k}\,|\vec{k}^{\,2}}{(M^2_{\rm N} + \vec{k}^{\,2})^{2/2}} =
\frac{2}{3}\,\Bigg(\frac{\Lambda_{\rm D}}{M_{\rm N}}\Bigg)^{\!3} \ll
1.\eqno({\rm A}.41)
$$
The cut--off  $\Lambda_{\rm D}$ restricts 3--momenta of the virtual nucleon
fluctuations forming the physical deuteron [1,2]. Due to the uncertainty
relation $\Delta r\,\Lambda_{\rm D} \ge 1/2$ the spatial  region of virtual
nucleon fluctuations forming the physical deuteron is defined by $\Delta r
\ge 1.44\,{\rm fm}$. This agrees with the range of nuclear forces (NF)
caused by the one--pion exchange with the mass $M_{\pi} =135\,{\rm MeV}$:
$r_{\rm N F} = 1/M_{\pi} = 1.46\,{\rm fm}$ [14].

After the continuation of the results of the calculation of the structure
functions on--mass shell of the deuteron the contribution of ${\cal
J}^{\mu\nu}(k_{\rm D}, k_{\ell}; Q)$ can be neglected relative to the
contribution of ${\cal J}^{\alpha\mu\nu}(k_{\rm D}, k_{\ell}; Q)$. The
contribution of the structure function ${\cal J}^{\alpha\mu\nu}(k_{\rm D},
k_{\ell}; Q)$ does not depend on the mass of  virtual nucleons and
according to Ref.~[17] can be valued as the anomaly of the AAV one--nucleon
triangle diagram. The ambiguity of the calculation of ${\cal
J}^{\alpha\mu\nu}(k_{\rm D}, k_{\ell}; Q)$ caused by the dependence on an
arbitrary shift of a virtual momentum can be fixed by requirement of gauge
invariance of the amplitude of the process p + p $\to$ D + e$^+$ +
$\nu_{\rm e}$ under gauge transformations of the deuteron field
$e^{*\mu}(k_{\rm D}) \to e^{*\mu}(k_{\rm D}) + \lambda\,k^{\mu}_{\rm D }$,
where $\lambda$ is an arbitrary parameter. This gives  $a = -1/2$ and the
structure function in the form [2]:
$$
{\cal J}^{\alpha\mu\nu}(k_{\rm D}, k_{\ell}; Q) = 3\,(k^{\alpha}_{\rm D}
g^{\nu\mu} - k^{\nu}_{\rm D} g^{\mu\alpha}).\eqno({\rm A}.42)
$$
The attraction of requirement of gauge invariance in order to remove
ambiguities of the structure function ${\cal J}^{\alpha\mu\nu}(k_{\rm D},
k_{\ell}; Q)$ and to fix the contribution of the anomaly is in complete
agreement with the derivation of the Adler--Bell--Jackiw axial anomaly
performed in terms of one--fermion loop diagrams [16].

Since we strive to draw a similarity between the RFMD and effective quark
models motivated by QCD applied to the derivation of Effective Chiral
Lagrangians, requirement of gauge invariance under gauge transformations of
the deuteron field
$$
D_{\mu}(x) \to D_{\mu}(x) + \partial_{\mu}f(x),\eqno({\rm A}.43)
$$
where $f(x)$ is a gauge function, can be justified by referring to a
dynamics of vector meson fields in these effective quark models [34--36].
The effective Lagrangian of the physical deuteron field $D_{\mu}(x)$, which
we apply to the calculation of one--nucleon loop diagrams describing
effective low--energy interactions of the deuteron coupled to nucleons and
other particles, reads [1,2]:
$$
{\cal L}(x) = - \frac{1}{2}\,D^{\dagger}_{\mu\nu}(x) D^{\mu\nu}(x) +
M^2_{\rm D}D^{\dagger}_{\mu}(x) D^{\mu}(x) - ig_{\rm
V}[\bar{p}(x)\gamma^{\mu}n^c(x) - \bar{n}(x)\gamma^{\mu}p^c(x)]\,D_{\mu}(x)
$$
$$
-ig_{\rm V}[\bar{p^c}(x)\gamma^{\mu}n(x) -
\bar{n^c}(x)\gamma^{\mu}p(x)]\,D^{\dagger}_{\mu}(x) +
\bar{p}(x)(i\gamma^{\mu}\partial_{\mu} - M_{\rm N}) p(x) +
\bar{n}(x)(i\gamma^{\mu}\partial_{\mu} - M_{\rm N}) n(x), \eqno({\rm A}.44)
$$
where $D_{\mu\nu}(x) = \partial_{\mu} D_{\nu}(x) -
\partial_{\nu}D_{\mu}(x)$. Since the deuteron field $D_{\mu}(x)$ couples to
the conserved nucleon current $ J^{\mu}(x) = - i g_{\rm V}\, [\bar{pn}(x)
\gamma^{\mu} n^c(x) - \bar{n}(x) \gamma^{\mu} p^c(x)]$, i.e.,
$\partial_{\mu}J^{\mu}(x) = 0$, invariance of the effective Lagrangian
Eq.~({\rm A}.44) under gauge transformations of the deuteron field Eq.({\rm
A}.43) is violated only by the mass term. The same problem encounters
itself for description of dynamics of vector meson fields ($\rho(770)$,
$\omega(780)$ and so on) in effective quark models with chiral $U(3)\times
U(3)$ symmetry [34--36]. Since these mesons are massive, the kinetic
Lagrangians of vector meson fields are not invariant under gauge
transformations of these fields. Nevertheless, for the derivation of
effective low--energy interactions of vector mesons coupled to other
particles requirement of gauge invariance turns out to be very important.
For example, by virtue of requirement of gauge invariance under gauge
transformations of $\rho(770)$ and $\omega(780)$ meson fields one can fix
unambiguously the coupling constant of the $\omega \rho\pi$ interaction
defined by the Adler--Bell--Jackiw axial anomaly [16] that plays an
important role for the correct description of the $\omega \to
\pi^+\pi^-\pi^0$ decay and many other low--energy processes.

Substituting the structure function Eq.~({\rm A}.42) in Eq.~({\rm A}.37) we
obtain the matrix element of the solar proton burning in the following
relativistically invariant form
$$
i{\cal M}({\rm p} + {\rm p} \to {\rm D} + {\rm e}^+ + \nu_{e}) =-
\,C(\eta)\,G_{\rm V}\,g_{\rm A}G_{\rm \pi NN}\frac{3g_{\rm V}}{8\pi^2}
$$
$$
\times\,(k^{\alpha}_{\rm D} g^{\nu\mu} - k^{\nu}_{\rm D} g^{\mu\alpha})
[\bar{u^c}(p_2)\gamma_{\alpha}\gamma^5 u(p_1)] [\bar{u}(k_{\nu_{\rm e}})
\gamma_{\nu}(1-\gamma^5) v(k_{\rm e^+})]\,e^*_{\mu}(k_{\rm D}).\eqno({\rm
A}.45)
$$
In the low--energy limit due to the low--energy reduction
$$
[\bar{u^c}(p_2)\gamma_{\alpha}\gamma^5 u(p_1)] \to - g_{\alpha
0}[\bar{u^c}(p_2) \gamma^5 u(p_1)] \eqno({\rm A}.46)
$$
the matrix element of the solar proton burning can be brought up to the form
$$
i{\cal M}({\rm p} + {\rm p} \to {\rm D} + {\rm e}^+ + \nu_{e}) =
C(\eta)\,G_{\rm V}\,g_{\rm A}\,M_{\rm N} G_{\rm \pi NN}\frac{3g_{\rm
V}}{4\pi^2}
$$
$$
\times\,[\bar{u^c}(p_2)\ \gamma^5 u(p_1)] [\bar{u}(k_{\nu_{\rm e}})
\gamma^{\mu}(1-\gamma^5) v(k_{\rm e^+})]\, e^*_{\mu}(k_{\rm D}), \eqno({\rm
A}.47)
$$
where we have set $k^0_{\rm D} = M_{\rm D} \simeq 2 M_{\rm N}$ valid
on--mass shell of the deuteron. This completes the calculation of the
matrix element of the solar proton burning.

Omitting the Gamow penetration factor $C(\eta)$ the residual part of the
matrix element of the solar proton burning  Eq.~({\rm A}.45)  can be
defined by the effective Lagrangian
$$
{\cal L}_{\rm pp\to D e^+ \nu_{\rm e}}(x) = g_{\rm A}G_{\rm \pi
NN}\frac{G_{\rm V}}{\sqrt{2}}\frac{3g_{\rm
V}}{8\pi^2}\,D^{\dagger}_{\mu\nu}(x)\,[\bar{p^c}(x)\gamma^{\mu}\gamma^5
p(x)]\,[\bar{\psi}_{\nu_{\rm e}}(x)\gamma^{\nu}(1 - \gamma^5) \psi_{\rm
e}(x)].\eqno({\rm A}.48)
$$
This Lagrangian is {\it local} in accordance with {\it the condition of
microscopic causality} [32].

The calculation of the matrix element of the disintegration of the deuteron
by anti--neutrinos  $\bar{\nu}_{\rm e}$ + D $\to$ e$^+$ + n + n can be
carried out by an analogous way and defined by the same structure
functions. The effective Lagrangian describing the matrix element of the
transition $\bar{\nu}_{\rm e}$ + D $\to$ e$^+$ + n + n can be written as
follows
$$
{\cal L}_{\rm \bar{\nu}_{\rm e}D \to e^+ nn }(x) = g_{\rm A}G_{\rm \pi
NN}\frac{G_{\rm V}}{\sqrt{2}}\frac{3g_{\rm
V}}{8\pi^2}\,D_{\mu\nu}(x)\,[\bar{n}(x)\gamma^{\mu}\gamma^5
n^c(x)]\,[\bar{\psi}_{\nu_{\rm e}}(x)\gamma^{\nu}(1 - \gamma^5) \psi_{\rm
e}(x)].\eqno({\rm A}.49)
$$
The effective Lagrangians Eq.~({\rm A}.48) and  Eq.~({\rm A}.49) testify
distinctly that the processes of the solar proton burning and the
disintegration of the deuteron by anti--neutrinos are governed by the same
dynamics of strong low--energy nuclear interactions in agreement with
charge independence of the weak interaction strength.

\section*{Appendix B. Low--energy elastic {\rm NN} scattering in the RFMD}

In this Appendix we show how in the RFMD one can describe a
phenomenological amplitude of elastic NN scattering by using the effective
four--nucleon interaction Eq.~(\ref{label1.1}). For simplicity we suggest
to consider the elastic low--energy np scattering. In the RFMD the
amplitude of the elastic low--energy np scattering can be written as follows
$$
{\cal M}({\rm n p} \to {\rm n p})(k) = - {\cal A}(k)\,\frac{4\pi}{M_{\rm
N}}\,[\bar{u}(p^{\prime}_2)\gamma^5
u^c(p^{\prime}_1)]\,[\bar{u^c}(p_1)\gamma^5 u(p_2)],\eqno ({\rm B}.1)
$$
since at low energies the interaction $\gamma^{\mu}\gamma^5\otimes
\gamma_{\mu}\gamma^5$ reduces to $\gamma^5\otimes \gamma^5$, then $p_i$ and
$p^{\prime}_i$ (i=1,2) are 4--momenta of the proton and the neutron in the
initial and final states and $k$ is a relative 3--momentum of the np
system. The phenomenological amplitude of the low--energy elastic np
scattering ${\cal A}(k)_{\rm ph}$ reads [12,14]
$$
{\cal A}(k)_{\rm ph} = \frac{1}{\displaystyle -\frac{1}{a_{\rm np}} +
\frac{1}{2}\,r_{\rm np}k^2 - i\,k} , \eqno ({\rm B}.2)
$$
where $a_{\rm np}$ and $r_{\rm np} = (2.75 \pm 0.05)\,{\rm fm}$ are the
S--wave scattering length and the effective range of the np--scattering in
the ${^1}{\rm S}_0$--state [10]. At $k\to 0$ we get ${\cal A}(0)_{\rm ph} =
- a_{\rm np}$ which gives the cross section equal $\sigma({\rm n p} \to
{\rm n p}) = 4 \pi a^2_{\rm np}$.

In the RFMD due to the low--energy reduction
$$
[\bar{n}(x)\gamma_{\alpha}\gamma^5
p^c(x)]\,[\bar{p^c}(x)\gamma^{\alpha}\gamma^5 n(x)] \to - [\bar{n}(x)
\gamma^5 p^c(x)]\,[\bar{p^c}(x)\ \gamma^5 n(x)] \eqno({\rm B}.3)
$$
the np scattering runs through the one--nucleon loop exchange. Using the
effective interaction Eq.~(\ref{label1.1}) we can write down the effective
Lagrangian for the low--energy elastic np scattering:
$$
\int d^4x\,{\cal L}^{\rm np \to np}_{\rm eff}(x)_{\rm scattering} = -
\frac{G^2_{\rm \pi NN}}{16\pi^2}\int d^4x\int \frac{d^4x_1
d^4k_1}{(2\pi)^4}\,e^{\textstyle - ik_1\cdot (x - x_1)}
$$
$$
\times\,\{[\bar{n}(x)\gamma_{\alpha}\gamma^5
p^c(x)][\bar{p^c}(x_1)\gamma_{\beta}\gamma^5 n(x_1)]\,{\cal
J}^{\alpha\beta}(k_1) + [\bar{n}(x)\gamma^5 p^c(x)][\bar{p^c}(x_1)\gamma^5
n(x_1)]\,{\cal J}(k_1)
$$
$$
+[\bar{n}(x)\gamma_{\alpha}\gamma^5 p^c(x)][\bar{p^c}(x_1)\gamma^5
n(x_1)]\,{\cal J}^{\alpha}(k_1) + [\bar{n}(x)\gamma^5
p^c(x)][\bar{p^c}(x_1)\gamma_{\alpha}\gamma^5 n(x_1)]\,{\bar{\cal
J}}^{\alpha}(k_1)\},\eqno ({\rm B}.4)
$$
where ${\cal J}^{\alpha\beta}(k_1)$, ${\cal J}(k_1)$, ${\cal
J}^{\alpha}(k_1)$ and ${\bar{\cal J}}^{\alpha}(k_1)$ are the structure
functions defined by the momentum integrals:
$$
{\cal J}^{\alpha\beta}(k_1) = \int \frac{d^4q}{\pi^2i}\,{\rm
tr}\Bigg\{\frac{1}{M_{\rm N} - \hat{q}}\gamma^{\alpha}\gamma^5
\frac{1}{M_{\rm N} - \hat{q} - \hat{k}_1}\gamma^{\beta}\gamma^5\Bigg\},
$$
$$
{\cal J}(k_1) = \int \frac{d^4q}{\pi^2i}\,{\rm tr}\Bigg\{\frac{1}{M_{\rm N}
- \hat{q}}\gamma^5 \frac{1}{M_{\rm N} - \hat{q} - \hat{k}_1}\gamma^5\Bigg\}
$$
$$
{\cal J}^{\alpha}(k_1) = \int \frac{d^4q}{\pi^2i}\,{\rm
tr}\Bigg\{\frac{1}{M_{\rm N} - \hat{q}}\gamma^{\alpha}\gamma^5
\frac{1}{M_{\rm N} - \hat{q} - \hat{k}_1}\gamma^5\Bigg\}
$$
$$
\bar{{\cal J}}^{\alpha}(k_1) = \int \frac{d^4q}{\pi^2i}\,{\rm
tr}\Bigg\{\frac{1}{M_{\rm N} - \hat{q}}\gamma^5 \frac{1}{M_{\rm N} -
\hat{q} - \hat{k}_1}\gamma^{\alpha}\gamma^5\Bigg\}.\eqno({\rm B}.5)
$$
The amplitude of the low--energy elastic np scattering defined by the
effective Lagrangian Eq.~({\rm B}.4) reads
$$
{\cal M}({\rm np \to np}) = - \frac{G^2_{\rm \pi NN}}{16\pi^2}
$$
$$
\times\,\{[\bar{u}(p^{\prime}_2)\gamma_{\alpha}\gamma^5
u^c(p^{\prime}_1)]\,[\bar{u^c}(p_1)\gamma_{\beta}\gamma^5 u(p_2)]\,{\cal
J}^{\alpha\beta}(P) + [\bar{u}(p^{\prime}_2)\gamma^5
u^c(p^{\prime}_1)]\,[\bar{u^c}(p_1)\gamma^5 u(p_2)]\,{\cal J}(P)
$$
$$
+[\bar{u}(p^{\prime}_2)\gamma_{\alpha}\gamma^5
u^c(p^{\prime}_1)]\,[\bar{u^c}(p_1)\gamma^5 u(p_2)]\,{\cal J}^{\alpha}(P) +
[\bar{u}(p^{\prime}_2)\gamma^5
u^c(p^{\prime}_1)]\,[\bar{u^c}(p_1)\gamma_{\alpha}\gamma^5
u(p_2)]\,{\bar{\cal J}}^{\alpha}(P)\} ,\eqno({\rm B}.6)
$$
where $P = p_1 + p_2 = p^{\prime}_1 + p^{\prime}_2$ and in the center of
mass frame $P^{\mu} = (2\sqrt{k^2 + M^2_{\rm N}}, \vec{0}\,)$.

Due to the low--energy reduction Eq.~({\rm B}.3) the amplitude Eq.~({\rm
B}.6) reduces to the form
$$
{\cal M}({\rm np \to np}) = - \frac{G^2_{\rm \pi NN}}{16\pi^2}
\,[\bar{u}(p^{\prime}_2)\gamma^5
u^c(p^{\prime}_1)]\,[\bar{u^c}(p_1)\gamma^5 u(p_2)]
$$
$$
\times\,[-{\cal J}^{00}(P) + {\cal J}(P) + {\cal J}^{0}(P) -\bar{{\cal
J}}^{0}(P)],\eqno({\rm B}.7)
$$
where the structure functions ${\cal J}^{00}(P)$, ${\cal J}(P)$, ${\cal
J}^{0}(P)$ and $\bar{{\cal J}}^{0}(P)$ are given by Eq.~({\rm B}.5) with
the change $k_1 \to P$.

The integrals over $q$ are both quadratically and logarithmically
divergent. In the RFMD they are regularized by a cut--off $\Lambda_{\rm D}
\ll M_{\rm N}$ [2]. Then, calculating the integrals over $q$ we have to
take into account that quadratically divergent integrals regularized by a
cut--off are defined ambiguously with respect to the shift of virtual
momenta. Indeed, it is well known [17,2] that
$$
\int \frac{d^4q}{\pi^2i}\,{\rm tr}\Bigg\{\frac{1}{M_{\rm N} - \hat{q} -
\hat{Q}}\gamma^{\alpha}\gamma^5 \frac{1}{M_{\rm N} - \hat{q} - \hat{Q} -
\hat{P}}\gamma^{\beta}\gamma^5\Bigg\} =
$$
$$
=\int \frac{d^4q}{\pi^2i}\,{\rm tr}\Bigg\{\frac{1}{M_{\rm N} -
\hat{q}}\gamma^{\alpha}\gamma^5 \frac{1}{M_{\rm N} - \hat{q} -
\hat{P}}\gamma^{\beta}\gamma^5\Bigg\} + 2\,[Q^{\alpha}(Q + P)^{\beta} +
Q^{\beta}(Q + P)^{\alpha} - Q\cdot (Q + P)\,g^{\alpha\beta}],
$$
$$
\int \frac{d^4q}{\pi^2i}\,{\rm tr}\Bigg\{\frac{1}{M_{\rm N} - \hat{q} -
\hat{Q}}\gamma^5 \frac{1}{M_{\rm N} - \hat{q} - \hat{Q} -
\hat{P}}\gamma^5\Bigg\} =
$$
$$
=\int \frac{d^4q}{\pi^2i}\,{\rm tr}\Bigg\{\frac{1}{M_{\rm N} - \hat{q}}
\gamma^5 \frac{1}{M_{\rm N} - \hat{q} - \hat{P}} \gamma^5\Bigg\} -
2\,Q\cdot (Q + P), \eqno({\rm B}.8)
$$
where a 4--momentum $Q$ defines an arbitrary shift of a virtual momentum $
q \to q + Q$. The most general form of $Q$ reads : $Q = \xi\,P + N$, where
$\xi$ is an arbitrary parameter and $N$ is an arbitrary 4--momentum
orthogonal to $P$, i.e., $P\cdot N = 0$. As $P$ is a time--like vector,
$P^2 > 0$, so  $N$ is a space--like, $N^2 < 0$. In the center of mass frame
we have $N^{\mu} = (0,\vec{N}\,)$. The result of the calculation of the
structure function $-{\cal J}^{00}(P) + {\cal J}(P)$ can be given in the
following general form:
$$
\frac{G^2_{\rm \pi NN}}{16\pi^2}\,[-{\cal J}^{00}(P) + {\cal J}(P)] =
\frac{4\pi}{M_{\rm N}}(C_1 + C_2\,k^2),\eqno({\rm B}.9)
$$
where $C_i$ ($i$=1,2) are arbitrary constants containing  all uncertainties
induced by shifts of virtual momenta. One should take into account that
virtual momenta in the momentum integrals defining the structure functions
${\cal J}^{\alpha\beta}(P)$ and ${\cal J}(P)$ can be shifted independently.
The structure functions ${\cal J}^{0}(P)$ and ${\bar{\cal J}}^{0}(P)$ are
logarithmically divergent and, therefore, do not depend on the shift of
virtual momentum. The contribution of these structure functions can be
absorbed by the constants $C_1$ and $C_2$:
$$
\frac{G^2_{\rm \pi NN}}{16\pi^2}\,[-{\cal J}^{00}(P) + {\cal J}(P) + {\cal
J}^{0}(P) - {\bar{\cal J}}^{0}(P)] = \frac{4\pi}{M_{\rm N}}(C_1 +
C_2\,k^2),\eqno({\rm B}.10)
$$
Inserting Eq.~({\rm B}.10) in the r.h.s. of Eq.~({\rm B}.7) we obtain the
amplitude of the low--energy elastic np scattering in terms of the
constants $C_1$ and $C_2$:
$$
{\cal M}({\rm np \to np}) = - \frac{4\pi}{M_{\rm N}}(C_1 + C_2\,k^2)
\,[\bar{u}(p^{\prime}_2)\gamma^5
u^c(p^{\prime}_1)]\,[\bar{u^c}(p_1)\gamma^5 u(p_2)].\eqno({\rm B}.11)
$$
Neglecting the terms of order $O(k^4)$ as it is accepted for the
description of low--energy elastic np scattering [12], we arrive at the
expression
$$
{\cal M}({\rm np \to np}) = - \frac{4\pi}{M_{\rm
N}}\,\frac{1}{\displaystyle \frac{1}{C_1} -
\frac{C_2}{C^2_1}\,k^2}\,[\bar{u}(p^{\prime}_2)\gamma^5
u^c(p^{\prime}_1)]\,[\bar{u^c}(p_1)\gamma^5 u(p_2)].\eqno({\rm B}.12)
$$
Matching Eq.~({\rm B}.12) with  Eq.~({\rm B}.1) we obtain ${\cal A}(k)_{\rm
RFMD}$ in the form:
$$
{\cal A}(k)_{\rm RFMD} = \frac{1}{\displaystyle \frac{1}{C_1} -
\frac{C_2}{C^2_1}\,k^2}. \eqno ({\rm B}.13)
$$
Following [12] we can set
$$
C_1 = -a_{\rm np}\quad, \quad C_2 = - \frac{1}{2}\,r_{\rm np}\,a^2_{\rm
np}.\eqno ({\rm B}.14)
$$
This yields the amplitude of the low--energy elastic np scattering in the form
$$
{\cal A}(k)_{\rm RFMD} = \frac{1}{\displaystyle -\frac{1}{a_{\rm np}} +
\frac{1}{2}\,r_{\rm np}k^2}. \eqno ({\rm B}.15)
$$
Up to the imaginary part $-i\,k$ which can be appended to the r.h.s. of
Eq.~({\rm B}.15) due to unitarity the amplitude
${\cal A}(k)_{\rm RFMD}$
coincides with the phenomenological amplitude
Eq.~({\rm B}.2). Thus,  we have shown that in the RFMD with the local
four--nucleon interaction Eq.~(\ref{label1.1}) one can describe, in spirit
of the EFT approach [11--13]  and in agreement with low--energy nuclear
phenomenology, the amplitude of the low--energy elastic np scattering in
terms of the S--wave scattering length $a_{\rm np}$ and the effective range
$r_{\rm np}$. This confutes the statement by Bahcall and Kamionkowski [43]
that the effective local four--nucleon interaction Eq.~(\ref{label1.1})
leads in the RFMD to the zero effective range for elastic NN scattering,
i.e., $r_{\rm NN} = 0$. This statement [43] is also confuted by our results
on the astrophysical factor for the solar proton burning, the reaction rate
for the neutron--proton radiative capture and the cross sections for  the
low--energy disintegration of the deuteron by photons and anti--neutrinos
obtained in good agreement with the PMA.

\end{document}